\documentclass[published]{JHEP3}

\JHEP{00(2007)000}

\JHEPspecialurl{http://jhep.sissa.it/JOURNAL/JHEP3.tar.gz}

\usepackage{epsfig,multicol,bbm}

\newcommand\fverb{\setbox\pippobox=\hbox\bgroup\verb}
\newcommand\fverbdo{\egroup\medskip\noindent%
            \fbox{\unhbox\pippobox}\ }
\newcommand\fverbit{\egroup\item[\fbox{\unhbox\pippobox}]}
\newbox\pippobox
\title{Gravity Dual of Superconformal Anomaly}

\author{W.F. Chen\\
 Physics Division, National Center for
Theoretical Sciences, Hsinchu, Taiwan, R.O.C. \\ and\\
  Department
of Physics, University of Winnipeg, Winnipeg, Manitoba, Canada, R3B
2E9\\
E-mail: \email{we.chen@uwinnipeg.ca}}

\received{xxx, 2007}        %%
\accepted{xxx, 2007}        %% These are for published papers.

\abstract{
 The supergravity dual of  superconformal anomaly in a
 four-dimensional supersymmetric gauge theory is investigated.
  We consider a well-established dual correspondence
  between the ${\cal N}=1$  $SU(N+M)\times SU(N)$ supersymmetric gauge
  theory with two flavors of  matter fields  in the bifundamental
  representation of gauge group and type IIB superstring in the
  space-time background furnished by the Klebanov-Strassler (K-S)
  solution. The $D$-brane configuration for these two dual theories
  consists of $N$ $D3$ branes and $M$ fractional $D3$ branes in
  the singular  space-time composed of a direct product of $M^4$
  and a six-dimensional conifold ${\cal C}_6$ with the base
  $T^{1,1}$. The superconformal anomaly originates from
  fractional branes frozen
at the apex of ${\cal C}_6$. While on the gravity side, the
fractional branes deform the $AdS_5\times T^{1,1}$ space-time
background and partially break local supersymmetry of type IIB
supergravity. We choose the K-S solution as the vacuum configuration
for type IIB supergravity and observe the five-dimensional gauged
supergravity produced from the spontaneous compactification on the
deformed $T^{1,1}$. Consequently, we find that the deformation on
$AdS_5\times T^{1,1}$ leads to the spontaneous breaking of
supersymmetry in gauged $AdS_5$ supergravity and a super-Higgs
mechanism arises. The graviton multiplet dual to the superconformal
current of supersymmetric gauge theory becomes massive with the
Goldstone multiplet relevant to the fluxes carried by fractional
branes. We thus conclude that the super-Higgs mechanism in gauged
supergravity is dual to the superconformal anomaly of supersymmetric
gauge theory in terms of gauge/gravity correspondence.}


\begin{document}

% \vspace{3ex}

\section{Introduction}

The AdS/CFT correspondence  and its  generalization, gauge/gravity
duality, describe a physical equivalence between a supersymmetric
gauge theory and a gravitational system \cite{mald,gkp,witt1,agmo}.
This amazing correspondence originates from open/closed string
duality bridged by D-brane, a solitonic object in string theory. A
D-bane has two distinct features \cite{polch}: on one hand, in
weakly coupled type-II superstring theory, it behaves as a dynamical
and geometrical object with open strings ending on it. Thus a stack
of coincided $Dp$-branes at low-energy will yield a
$p+1$-dimensional supersymmetric gauge theory on the world-volume of
$Dp$-branes. In principle, one can use various $D$-branes and NS
solitonic branes as well as  other geometric objects like
orientifold planes to build any possible brane configurations and
construct a anticipated gauge theory \cite{giku, direv}; On the
other hand, a $Dp$-brane carries R-R charge and couples with
$p+1$-rank antisymmetric field obtained from the R-R boundary
condition in closed string theory. So it provides a source to the
low-energy effective theory of strongly coupled type-II superstring
theory --- type-II supergravity. Therefore, a stack of $Dp$-branes
can modify space-time background of string theory and emerge as a
$p$-brane solution to type-II supergravity \cite{stel, bcpz}. This
is the essence for gauge/gravity duality.

 A physical phenomenon in one theory should
have a physically equivalent description in the dual theory in terms
of gauge/gravity duality. The quantum anomaly, a violation of
classical symmetry by quantum correction, is a typical quantum
phenomenon in field theory.  The investigation on the dual
description of anomaly can
 reveal some important  features in gauge/gravity duality. Not
long time ago, Klebanov, Ouyang and Witten investigated the gravity
dual of $U(1)_R$ $R$-symmetry anomaly of $N=1$
 cascading $SU(N+M)\times  SU(N)$ gauge
 theory \cite{kow} in terms of  its dual theory --- type IIB supergravity
 in the background furnished by K-S solution \cite{klst}. The K-S solution is
 non-singular and originates from the brane configuration consisting
 of $N$ bulk $D3$-branes and $M$ fractional $D3$-branes  fixed at the apex
 of conifold ${\cal C}_6$ in the target space-time $M_4\times {\cal
 C}^6$ \cite{klwi}.
 If the fractional branes are absent in the brane configuration,
 the resultant space-time background is  three-brane solution with
  near-horizon limit $AdS_5\times T^{1,1}$. The spontaneous
 compactification on $T^{1,1}$ of type IIB supergravity
 yields  ${\cal N}=2$ $U(1)$ gauged $AdS_5$ supergravity.
   The UV limit of K-S solution can be considered  a deformed $AdS_5\times T^{1,1}$.
 Klebanov {\it et al} showed that
 the $U(1)_R$ anomaly of ${\cal N}$=1 $SU(N+M)\times SU(N)$
supersymmetric gauge theory is dual to a spontaneous breaking of
$U(1)$ gauge symmetry in gauged $AdS_5$ supergravity
\cite{guna,gstow} and the consequent Higgs mechanism.

However, in a classical scale invariant ${\cal N}=1$ supersymmetric
gauge theory, the chiral $R$-symmetry current $j^\mu$, the
supersymmetry current $s^\mu$ and energy-momentum tensor
$\theta^{\mu\nu}$, ($j^\mu$,$s^\mu$,$\theta^{\mu\nu}$), constitute a
current supermultiplet \cite{ilio,anm}.  At quantum level
  the scale symmetry,  chiral $R$-symmetry and  conformal
  supersymmetry in an ${\cal
N}<4$ supersymmetric gauge theory become anomalous due to the
non-vanishing beta function of the
  theory. The divergence $\partial_\mu j^\mu$ of chiral $R$-symmetry
   current $j^\mu$, the $\gamma$-trace $\gamma_\mu s^\mu$ of supersymmetry
   current $s^\mu$ and  the trace $\theta^\mu_{~\mu}$ of energy-momentum
   tensor $\theta_{\mu\nu}$ also constitute an anomaly supermultiplet
   with respect to the Poincar\'{e}
   supersymmetry \cite{grisa}.  Therefore, one may ask
   what the dual of the whole superconformal anomaly is. A natural
   guess is that it should correspond to the spontaneous breaking of
   local supersymmetry and the consequent super-Higgs
   mechanism in gauged $AdS_5$ supergravity. The aim of this
   paper is to show  quantitatively how this
   dual description  to the superconformal anomaly indeed stands there.

In Sect.\,\ref{sect2}, we first use the ${\cal N}=1$ supersymmetric
Yang-Mills theory to illustrate the superconformal current and its
anomaly supermultiplet. To observe the origin of the superconformal
anomaly in the brane configuration, we show how superconformal
anomaly is reflected in the local part of gauge invariant quantum
effective action. Further, we introduce the
 ${\cal N}=1$ $SU(N+M)\times SU(N)$ supersymmetric gauge theory
with two chiral matter superfields $A_i$, $B_i$ ($i=1,2$) in the
bifundamental representation of gauge group and its superconformal
anomaly.
 Sect.\,\ref{sect3} displays the origin of superconformal anomaly in
 $D$-brane configuration.
We review the $D3$/fractional $D3$ brane system in the singular
space-time $M^4\times {\cal C}_6$. This brane configuration leads to
${\cal N}=1$ $SU(N+M)\times SU(N)$ supersymmetric gauge theory. By
expanding the Dirac-Born-Infeld (DBI) action and Wess-Zumino (WZ)
term for  a $D$-brane configuration and comparing with the quantum
effective action of the supersymmetric gauge theory, we demonstrate
that the superconformal anomaly originates from the fractional brane
locating in the singularity of background space-time. We  further
introduce the K-S solution and its UV limit for later use. In
Sect.\,\ref{sect4}, we analyze the manifestation of various
symmetries on K-S solution. By comparing with three-brane solution
without fractional branes, we show how the presence of fractional
branes breaks some of local symmetries. We emphasize that it is the
NS-NS and R-R two-form fields
 relevant to fractional branes that break $U(1)$ rotation-
 and scale transformation invariance in the transverse space
 of the above  $D$-brane configuration. These two geometrical symmetries
 in the brane configuration are  the global $U(1)_R$-
 and scale symmetry of the supersymmetric
 gauge theory living on the world-volume of $D3$-branes. Further,
 the Killing spinor equation in the
  K-S solution background
 tells  how half of supersymmetries disappear due to
 the modification on $AdS_5\times T^{1,1}$  by fractional branes.
In Sect.\,\ref{sect5} we choose the K-S solution as a vacuum
configuration for type IIB supergravity and observe the dynamical
phenomenon. The spontaneous compactification on the deformed
$T^{1,1}$ occurs and a five-dimensional gauged supergravity arises.
In contrast to the case in the $AdS_5\times T^{1,1}$ background,
which gives the ${\cal N}=2$ $U(1)$ gauged $AdS_5$ supergravity
coupled with $SU(2)\times SU(2)$ Yang-Mills vector multiplets as
well as some Betti scalar, vector and tensor multiplets, we find
that the local supersymmetry, diffeomorphism symmetry  and $U(1)$
gauge symmetry  partially break in the gauged $AdS_5$ supergravity
since the deformed $AdS_5\times T^{1,1}$ possesses less isometric
symmetries and preserves less supersymmetries. We then go to reveal
the super-Higgs mechanism corresponding to the spontaneous breaking
of the above local symmetries. We exhibit the process  how the
graviton, $U(1)$ gauge field and gravitino in $AdS_5$ space become
massive through ``eating" a Goldstone supermutiplet originating from
the NS-NS- and R-R fluxes carried by fractional branes. Since the
deformation on $AdS_5\times T^{1,1}$ is produced by fractional
branes in the brane configuration and meanwhile they are the origin
for superconformal anomaly on the field theory side,
 we thus claim that the gravity
dual of superconformal anomaly is the spontaneous breaking of local
supersymmetry and the consequent super-Higgs effect in gauged
$AdS_5$ supergravity. Sect.\,\ref{sect6} is a brief summary.

\section{Superconformal anomaly multiplet  of ${\cal N}=1$
  four-dimensional supersymmetric gauge theory}
\label{sect2}

\subsection{${\cal N}=1$ supersymmetric $SU(N)$ Yang-Mills
theory as an illustration}  \label{subsect21}

In a an ${\cal N}=1$  supersymmetric  field theory, the Poincar\'{e}
supersymmetry combines the energy-momentum tensor $\theta^{\mu\nu}$,
the supersymmetry current $s^{\mu}$ and the axial vector (or
equivalently chiral) R-symmetry current $j^{\mu}$   into a
supercurrent multiplet. If a field model has scale symmetry, these
currents are not only conserved at classical level,
\begin{eqnarray}
\partial_\mu \theta^{\mu\nu}=\partial_\mu s^\mu=\partial_\mu
j^{\mu}=0,
\end{eqnarray}
but also satisfy  algebraic relations,
$\theta^{\mu}_{~\mu}=\gamma_\mu s^\mu=0$. Consequently, three more
conserved currents can be constructed,
\begin{eqnarray}
d^\mu {\equiv} x_\nu \theta^{\nu\mu}, ~~~ k_{\mu\nu}{\equiv} 2 x_\nu
x^\rho\theta_{\rho\mu}-x^2\theta_{\mu\nu}, ~~~ l_\mu{\equiv} ix^\nu
\gamma_\nu s_\mu.
\end{eqnarray}
The charges  $D$, $K_\mu$ and  $S_\alpha$ of these three new
currents are the generators for dilatation, special conformal
symmetry- and conformal supersymmetry transformations. They
constitute an ${\cal N}=1$ superconformal algebra $SU(2,2|1)$
together with Lorentz generators $M_{\mu\nu}$, translation generator
$P_\mu$ and supercharge $Q_\alpha$. Consequently, the Poincar\'{e}
supersymmetry promotes to a larger superconformal symmetry.

 However,  the superconformal symmetry  becomes
 anomalous at quantum level. For an ${\cal N} <4 $
 supersymmetric gauge theory, due to the renormalization
 effect, an energy   scale is be generated dynamically.
Consequently,  the scale anomaly and the subsequent superconformal
anomaly multiplet required by the Poincar\'{e} supersymmetry
 must arise.
 In the case that all of them, the trace $\theta^\mu_{~\mu}$ of
 energy-momentum tensor $\theta_{\mu\nu}$, the $\gamma$-trace $\gamma^\mu s_\mu$
 of supersymmetry current $s_\mu$  and the divergence  $\partial_\mu j^{\mu}$
 of the chiral $R$-symmetry current $j_\mu$,  get
 contribution from quantum correction,
$\left(\partial_\mu j^{\mu},\gamma^\mu s_\mu,
\theta^{\mu}_{~\mu}\right)$
  will form a  chiral supermultiplet with  $\partial_\mu
j^{\mu}$ playing the role of the lowest component of a chiral
superfield \cite{ilio,anm}.

 In the following we take
${\cal N}=1$ supersymmetric $SU(N)$ Yang-Mills theory as an
illustrating example. Its field content consists of a vector field
$A_\mu$, a Majorana spinor $\lambda$ and an auxiliary field $D$ in
the adjoint representation of $SU(N)$. The classical action of the
theory is
\begin{eqnarray}
S=\int d^4x \left[-\frac{1}{4}G_{\mu\nu}^a G^{\mu\nu a}+\frac{1}{2}
i\overline{\lambda}^a\gamma^\mu (\nabla_\mu \lambda)^a + \frac{1}{2}
\left(D^a\right)^2\right].
\end{eqnarray}
The superconformal anomaly multiplet at quantum level is
\cite{grisa}
\begin{eqnarray}
\partial_\mu j^{\mu} &=& {2}\frac{\beta (g)}{g}
\left[-\frac{1}{6}G^a_{\mu\nu}\widetilde{G}^{a\mu\nu}+\frac{1}{3}
\partial^\mu \left( \overline{\lambda}^a\gamma_\mu\gamma_5\lambda^a\right)
\right],\nonumber\\
 \gamma^\mu s_\mu &=& 2\frac{\beta
(g)}{g}\left(-\sigma^{\mu\nu}G^a_{\mu\nu}+\gamma_5
D^a\right)\lambda^a,
\nonumber\\
\theta^\mu_{~\mu} &=& 2\frac{\beta
(g)}{g}\left[-\frac{1}{4}G_{\mu\nu}^a G^{\mu\nu a}+\frac{1}{2}
i\overline{\lambda}^a\gamma^\mu (\nabla_\mu \lambda)^a+\frac{1}{2}
\left(D^a\right)^2\right], \label{sanomaly1}
\end{eqnarray}
In above equation, $\beta (g)$ is the $\beta$-function for the gauge
coupling of the ${\cal N}=1$ $SU(N)$ super-Yang-Mills theory. At
one-loop order, $\beta (g)=-{3Ng^3}/(16\pi^2)$.

 Eq.\,(\ref{sanomaly1}) shows that the
anomaly coefficients are proportional to the beta function of ${\cal
N}=1$ supersymmetric Yang-Mills theory \cite{grisa,salam,abb}. This
is a universal feature for any ${\cal N} <4$ supersymmetric field
theory. The origins of scale anomaly $\theta^\mu_{~\mu}$ and chiral
anomaly $\partial_\mu j^\mu$ are the same as in usual gauge field
theories: the UV divergence in perturbation theory requires a
renormalization procedure to make the theory well defined in which
an energy scale parameter must be introduced, hence the scale
anomaly arises. The origin of chiral $U_R(1)$ anomaly is similar to
that of the axial $U_A(1)$ anomaly in QCD. It comes from the
contradiction of $U_R(1)$ symmetry with the global vector $U(1)$
rotational symmetry among gauginoes at quantum level. The $U_R(1)$
symmetry thus becomes anomalous when the vector $U(1)$ symmetry is
required to stand. The outcome of the $\gamma$-trace anomaly lies in
the incompatibility between the conservation  and the vanishing
$\gamma$-trace of supersymmetry current $s^\mu$ at quantum level
\cite{abb}. Usually, from the physical consideration that the
supersymmetry current should be conserved, the $\gamma$-trace
anomaly thus takes place.

 The superconformal anomaly manifest itself in the quantum effective action of
the theory. We first re-scale the fields
 $(A_\mu,\lambda,D)\rightarrow (A_\mu/g, \lambda/g,D/g)$
 and consider the strong CP violation term. Then
 the classical action of ${\cal N}=1$ supersymmetric $SU(N)$
 Yang-Mills theory can be re-written as
\begin{eqnarray}
 S_{\rm cl}= \frac{1}{g^2}\int d^4x\left[-\frac{1}{4}G_{\mu\nu}^a G^{\mu\nu
a}+\frac{1}{2} i\overline{\lambda}^a\gamma^\mu (\nabla_\mu
\lambda)^a + \frac{1}{2} \left(D^a\right)^2\right]
 +\frac{\theta^{\rm (CP)}}{32\pi^2}\int d^4x
 G_{\mu\nu}^a\widetilde{G}^{\mu\nu a}.
 \label{claco}
\end{eqnarray}
The local part of the gauge invariant quantum effective action takes
the form,
\begin{eqnarray}
\Gamma_{\rm eff}&=&\frac{1}{g^2_{\rm eff}}\int
d^4x\left[-\frac{1}{4}G_{\mu\nu}^a G^{\mu\nu a}+\frac{1}{2}
i\overline{\lambda}^a\gamma^\mu (\nabla_\mu \lambda)^a + \frac{1}{2}
\left(D^a\right)^2\right]\nonumber\\
&& +\frac{\theta^{\rm (CP)}_{\rm eff}}{32\pi^2}\int d^4x
 G_{\mu\nu}^a\widetilde{G}^{\mu\nu a},
 \label{quanac}
\end{eqnarray}
where all the fields are renormalized quantities. The scale- and
chiral anomalies are reflected in the running of gauge coupling and
the shift of $\theta$-angle due to the non-vanishing $\beta (g)$
\cite{pequ},
\begin{eqnarray}
\frac{1}{g^2_{\rm
eff}(q^2)}&=&\frac{1}{g^2(q_0^2)}+\frac{3N}{8\pi^2}\ln
\frac{q_0}{q} \equiv
\frac{ 3N}{8\pi^2}\ln \frac{q}{\Lambda}, \nonumber\\
\theta^{\rm (CP)}_{\rm eff} &=& \theta^{\rm (CP)}+3N.
\label{effcou}
\end{eqnarray}

 To recognize  the $\gamma$-trace
anomaly of supersymmetry current in the quantum effective action, we
must resort to its superfield form. First, the superfield form of
the classical Lagrangian (\ref{claco}) is
\begin{eqnarray}
{\cal L}=\frac{1}{32\pi} \int d^2\theta\, \mbox{Im}\left[\tau
\mbox{Tr} \left(W^\alpha W_\alpha\right) \right] +\mbox{h.c.}.
\end{eqnarray}
In above equation, $W_\alpha$ is the field strength of superspace
gauge connection and in the Wess-Zumino gauge, $W_\alpha
=\lambda_\alpha-\frac{i}{2}\left(\sigma^{\mu\nu}
\right)_{\alpha\beta}\theta^\beta F_{\mu\nu}+i\theta_\alpha
D-i\theta^2\left(\sigma^\mu\right)_{\alpha\dot{\alpha}}
 \nabla \overline{\lambda}^{\dot{\alpha}}$.
The parameter $\tau$ is a complex combination of gauge coupling and
$\theta$-angle,
\begin{eqnarray}
\tau=\frac{\theta^{\rm (CP)}}{2\pi} +i\frac{4\pi}{g^2}.
\end{eqnarray}
To show how the conformal supersymmetry anomaly $\gamma_\mu s^\mu$
is reflected in the quantum effective action, we define that $\tau$
is the lowest component of a certain constant chiral superfield,
$\Sigma \equiv \tau+\theta \chi +\theta^2 d$. Further, we require
that classically  $\chi_{\rm cl}=d_{\rm cl}=0$ and assume only at
quantum level $\chi$ and $d$ get non-vanishing expectation values,
\begin{eqnarray}
\Sigma_{\rm eff} &\equiv& \tau_{\rm eff}+\theta \chi_{\rm eff}
+\theta^2 d_{\rm eff}, \label{supercou}
\end{eqnarray}
where $\theta^{\rm (CP)}_{\rm eff}$, $g^2_{\rm eff}$ and hence
$\tau_{\rm eff}$ are given in (\ref{effcou}).
 In the Wess-Zumino gauge, the variation of the local quantum effective action
 in superspace produced from the shift of $\Sigma_{\rm eff}$ reads,
\begin{eqnarray}
\delta{\Gamma}_{\rm eff}&=& \frac{1}{32\pi}\int d^2\theta\,
\mbox{Im}\left[\delta\Sigma_{\rm eff} \mbox{Tr} \left(W^\alpha
W_\alpha\right) \right]+\mbox{h.c.}
 \nonumber\\
&=&\frac{1}{32\pi}\int d^2\theta\,
\mbox{Im}\left\{\left(\delta\tau_{\rm eff} +\theta \delta\chi_{\rm
eff} +\theta^2 \delta d_{\rm eff} \right) \left[ \lambda\lambda
+i\theta \left(2\lambda D+\sigma^{\mu\nu}\lambda G_{\mu\nu}\right)
\right.\right.\nonumber\\
&&\left.\left. +\theta^2 \left(-\frac{1}{2}G_{\mu\nu}G^{\mu\nu}+
\frac{i}{2}G_{\mu\nu}\widetilde{G}^{\mu\nu}-2i \lambda \sigma^\mu
\nabla_\mu \overline{\lambda} -D^2 \right)\right]
\right\}+\mbox{h.c.}. \label{superac}
\end{eqnarray}
 A comparison between  Eq.\,(\ref{superac}) and the superconformal
anomaly equation (\ref{sanomaly1}) reveals that the $\gamma$-trace
anomaly of supersymmetry current is reflected in the shift of the
fermionic parameter $\chi$, which is exactly like the trace- and
chiral anomalies are represented by the running of the gauge
couplings and the shift of $\theta$-angle. Note that this is not the
whole advantage of promoting $\tau$ parameter to a chiral
superfield. Later we shall later that from the viewpoint of brane
dynamics, the running of gauge coupling and the shift of $\theta$
angle originate from fractional branes locating at the space-time
singularity and further can get down to the Goldstone fields.
Therefore, introducing an artificial superpartner for the gauge
coupling and $\theta$-angle
 is helpful for identifying the Goldstone supermultiplet
 for the super-Higgs mechanism on
 the gauged $AdS_5$ supergravity side.

\subsection{${\cal N}=1$ supersymmetric $SU(N+M)\times SU(N)$
gauge theory with two flavors in bifundamental  representations
$(N+M,\overline{N})$ and $( \overline{N+M},N)$} \label{subsect22}

The reason why we specialize to ${\cal N}=1$ supersymmetric
$SU(N+M)\times SU(N)$ gauge theory with two flavors in the
bifundmental representations  $(N+M,\overline{N})$ and $(
\overline{N+M},N)$ of gauge group is that its supergravity dual is
fully understood, which is the type IIB superstring in the
space-time background found by Klebanov and Strassler \cite{klst}.
We first write down the classical action of this supersymmetric
gauge theory in superspace,
\begin{eqnarray}
&& S= \int d^4x\left\{\frac{1}{32\pi}\int  d^2\theta
\sum_{i=1}^2\mbox{Im}\left[\tau_{(i)} \mbox{Tr}\left(W^{(i)\alpha}
W^{(i)}_\alpha \right)\right]+\mbox{h.c.}\right. \nonumber\\
&+&\left. \int d^2\theta d^2\overline{\theta}\sum_{j=1}^2\left(
\overline{A}^j e^{ \left[g_1V^{(1)} (N+M)+g_2
V_2(\overline{N})\right]} A_j +\overline{B}_j e^{\left[g_1V^{(1)}
(\overline{N+M})+g_2 V_2(N)\right]}B^j\right)\right\}
\end{eqnarray}
where the two flavors $A_j$ and $B^j$ are chiral superfields in the
bifundamental representations $(N+M,\overline{N})$ and $(
\overline{N+M},N)$, respectively,
\begin{eqnarray}
\left(A_j\right)_{~r}^{m} &=& \left(\phi_{A j}\right)_{~r}^{m}
+\theta \left(\psi_{Aj}\right)_{~r}^{m}+\theta^2
\left(F_{Aj}\right)_{~r}^{m},\nonumber\\
 \left(B^j\right)^r_{~m}
&=&\left(\phi_{B}^{j}\right)^r_{~m}+\theta
\left(\psi_{B}^{j}\right)^r_{~m}+\theta^2
\left(F_{B}^{j}\right)^r_{~m},
\nonumber\\
 j&=& 1,2, ~~~ r=1,\cdots, N+M, ~~~ m=1,\cdots,N.
\end{eqnarray}
This model also admits a quartic superpotential,
%\begin{eqnarray}
$W=\lambda\epsilon^{ik}\epsilon^{jl}\mbox{Tr}\left(A_iB_jA_k B_l
\right)$. Since the superpotential should have $R$-charge $2$,  this
quartic superpotential normalizes  the R-charge of the chiral
superfields to be $1/2$. Consequently, their fermionic components
have $R$-charge $-1/2$.
 This model is a vector supersymmetric gauge theory. If we
consider $A_j$ as  quark chiral superfields, then $B^j$ are the
corresponding anti-quark chiral superfields. In the Wess-Zumino
gauge, the two-component form of above action is
\begin{eqnarray}
{\cal
L}&=&\sum_{i=1}^2\frac{1}{g_{(i)}^2}\left[-\frac{1}{4}G^{(i)a_i\mu\nu}
G^{(i)a_i}_{\mu\nu} +i\lambda^{(i)a_i}\sigma^{\mu}
\left(\nabla_{\mu}\overline{\lambda}^{(i)}\right)^{a_i}
+\frac{1}{2}D^{(i)a_i}D^{(i)a_i} \right] +\frac{\theta_{(i)}^{\rm
(CP)}}{32\pi^2}G^{(i)a_i\mu\nu} \widetilde{G}^{(i)a_i}_{\mu\nu}
\nonumber\\
 &+& \left(D^\mu \phi_A\right)^{\dagger j} D_\mu \phi_{Aj}+
\widetilde{D}^\mu \phi_B^j \left(\widetilde{D}_\mu
\phi_B\right)^\dagger_j
 + i\overline{\psi}^j_A \overline{\sigma}^\mu D_{\mu} \psi_{Aj}
+ i{\psi}_{B}^j{\sigma}^\mu\widetilde{D}_\mu \overline{\psi}_{Bj}
\nonumber\\
&+&  \sqrt{2}i \sum_{i=1}^2 (-1)^{i+1} g_{(i)}\left[ \phi^{\dagger
j}_A
 T^{a_i} \lambda^{a_i}_{(i)}\psi_{Aj}
 -\overline{\lambda}^{a_i}_{(i)}\overline{\psi}^j_A
 T^{a_i}\phi_{Aj}-
 \psi^j_B\lambda^{a_i}_{(i)} T^{a_i}{\phi}^\dagger_{Bj}+
 {\phi}^j_BT^{a_i}\overline{\psi}_{Bj}
 \overline{\lambda}_{(i)}^{a_i}\right] \nonumber\\
 &+& \sum_{i=1}^2 g_{(i)} (-1)^{i+1} D_{(i)}^{a_i}
 \left[\phi^{\dagger j}_A T^{a_i}\phi_{Aj}-{\phi}^j_BT^{a_i}
 {\phi}^\dagger_{Bj}\right]\nonumber\\
 &+&F^{\dagger j}_A F_{Aj}+{F}^j_B{F}^\dagger_{Bj}
 +\left[ \left(F_{Aj}\frac{\partial W}{\partial \phi_{Aj}}
 +F_{B}^j\frac{\partial W}{\partial \phi_{B}^j}\right)\right.
 \nonumber\\
 &+&\left.
 \frac{1}{2}\left(\frac{\partial^2 W}{\partial \phi_{Ai}
\partial \phi_{Aj}}\psi_{Ai} \psi_{Aj}
+\frac{\partial^2 W}{\partial \phi_{B}^i
\partial \phi_{B}^j}\psi_{B}^i \psi_{B}^j
 +2\frac{\partial^2 W}{\partial \phi_{Ai}
\partial \phi_B^j}\psi_{Ai} \psi_{B}^j
  \right)+\mbox{h.c.} \right],
  \label{lag}
\end{eqnarray}
where
\begin{eqnarray}
\left(\nabla_\mu \overline{\lambda}_{(i)}\right)^{a_i} &=&
\partial_\mu \overline{\lambda}_{(i)}^{a_i}+g
f^{a_ib_ic_i}A_{(i)\mu}^{b_i}
\overline{\lambda}_{(i)}^{c_i},\nonumber\\
\left(D_\mu \psi_{Aj}\right)^m_{~r}
 &=& \partial_\mu \psi_{Ajr}^m +ig_1A_{(1)\mu}^{a_1}
\left(T^{a_1}\right)_r^{~s}\psi_{Ajs}^m
-ig_2\psi_{Ajr}^n\left( T^{a_2}\right)_n^{~m} A_{(2)\mu}^{a_2},
\nonumber\\
\left(\widetilde{D}_\mu \overline{\psi}_{Bj}\right)_{~r}^m &=&
\partial_\mu \overline{\psi}_{Bjr}^m+
ig_1 \overline{\psi}_{Bjs}^m \left(T^{a_1}\right)^s_{~r}
A_{(1)\mu}^{a_1} -ig_2A_{(2)\mu}^{a_2}\left(T^{a_2}\right)^m_{~n}
\overline{\psi}_{Bjr}^n,\nonumber\\
\left(D_\mu\phi_{Aj}\right)_{~r}^m &=&
\partial_\mu\phi_{Ajr}^m+ig_1A_{(1)\mu}^{a_1}
\left(T^{a_1}\right)_r^{~s} \phi_{As}^m
-ig_2\phi_{Ajn} \left(T^{a_2}\right)^n_{~m}
 A_{(2)\mu}^{a_2} ,\nonumber\\
\left(D_\mu\phi_{A}^j\right)^{*r}_{~~m} &=& \partial_\mu
\phi^{*jr}_{Am} -ig_1\phi^{*js}_{Am} \left(T^{a_1}\right)_s^{~r}
A_{(1)\mu}^{a_1} + ig_2 A_{(2)\mu}^{a_2}\left( T^{a_2}\right)_m^{~n}
 \phi^{*jr}_{An},\nonumber\\
\left(\widetilde{D}_\mu {\phi}_B^j\right)^{r}_{~m} &=&
\partial_\mu \widetilde{\phi}_{Bm}^{jr}
-ig_1 {\phi}_{Bm}^s \left(T^{a_1}\right)_{s}^{~r}
 A_{(1)\mu}^{a_1}
+ig_2A_{(2)\mu}^{a_2} \left(T^{a_2}\right)_m^{~n}
 {\phi}_{Bn}^{jr},
\nonumber\\
\left(\widetilde{D}_\mu{\phi}_{Bj}\right)^{*m}_{~~r} &=&
\partial_\mu {\phi}^{*m}_{Bjr}
+ig_1A_{(1)\mu}^{a_1} \left(T^{a_1}\right)_r^{~s}
{\phi}^{*m}_{Bjs}-ig_2{\phi}^{*n}_{Bjs}
\left(T^{a_2}\right)_n^{~m} A_{(2)\mu}^{a_2}. \label{der}
\end{eqnarray}
The theory has global symmetry $SU_L(2)\times SU_R(2) \times
U_B(1)\times U_A(1)$  at classical level and the $U_A(1)$ symmetry
becomes anomalous at quantum level. In addition, the theory has a
chiral $U_R(1)$ R-symmetry which rotates the left- and right-handed
components  of ${\cal N}=1$ supercharges. It is this chiral $U_R(1)$
symmetry anomaly that enters the superconformal anomaly multiplet.

Let us list the supercurrent- and superconformal anomaly multiplets
of this field model to facilitate the analysis on superconformal
anomaly. The supercurrent multiplet (in four-component form) is
\begin{eqnarray}
\theta_{\mu\nu}&=&\sum_{i=1}^2\left\{-G_{\mu\rho}^{(i)a_i}
G_{~~\nu}^{(i)a_i\rho} +\frac{1}{2}i\left[\overline{\lambda}^{a_i}
(\gamma_\mu \nabla_\nu +\gamma_\nu \nabla_\mu)\lambda^{a_i}-
(\nabla_\mu \overline{\lambda}^{a_i}\gamma_\nu +\nabla_\nu
\overline{\lambda}^{a_i}\gamma_\mu)
\lambda^{a_i}\right]\right.\nonumber\\
&&\left.+\frac{1}{4}g_{\mu\nu} G_{\lambda\rho}^{(i)a_i}
G^{(i)a_i\lambda\rho}-\frac{1}{4}i g_{\mu\nu}
\left[\overline{\lambda}^{a_i} \gamma^\rho\left( \nabla_\rho
\lambda^{a_i}\right) -\left(\nabla_\rho\overline{\lambda}^{a_i}
\right) \gamma^\rho \lambda^{a_i}\right]\right\}
\nonumber\\
&&+\left[\left(D_\mu\phi_A\right)^{\dagger j}D_\nu \phi_{Aj}+
\left(D_\nu\phi_A\right)^{\dagger j}D_\mu \phi_{Aj}\right]+
\left[\left(\widetilde{D}_\mu\phi_B\right)^{\dagger}_j
\widetilde{D}_\nu \phi_{B}^j +
\left(\widetilde{D}_\nu\phi_B\right)^{\dagger}_j
\widetilde{D}_\mu\phi_{B}^j\right]
\nonumber\\
&&-\frac{1}{3}\left(\partial_\mu\partial_\nu-g_{\mu\nu}
\partial^2
\right)\left(\phi_A^{\dagger j}\phi_{Aj}\right)
-\frac{1}{3}\left(\partial_\mu\partial_\nu-g_{\mu\nu}\partial^2
\right)\left(\phi_{Bj}^{\dagger}\phi_{B}^j\right) \nonumber\\
&&+i\overline{\psi}^j\left(\gamma_\mu D_\nu+\gamma_\nu D_\mu
\right) \psi_j -g_{\mu\nu}\left\{i\overline{\psi}^j\gamma^\lambda
D_\lambda\psi_j +\left(D^\lambda \phi_A\right)^{\dagger j} D_\mu
\phi_{Aj}+ \widetilde{D}^\lambda \phi_B^j
\left(\widetilde{D}_\lambda \phi_B\right)^\dagger_j
\right.\nonumber\\
&&  + \frac{i} {\sqrt{2}} \sum_{i=1}^2 (-1)^{i+1} g_{(i)} \left[
\phi^{\dagger j}_A
 T^{a_i} \overline{\lambda}^{a_i}(1-\gamma_5)\psi_{j}
 -\overline{\psi}^j(1+\gamma_5){\lambda}^{a_i}
 T^{a_i}\phi_{Aj}\right.\nonumber\\
 && \left.-
 \overline{\psi}^j(1-\gamma_5)\lambda^{a_i}
 T^{a_i}{\phi}^\dagger_{Bj}
 +{\phi}^j_BT^{a_i}\overline{\lambda}^{a_i}(1+\gamma_5)
 {\psi}_{j}
 \right]-\frac{1}{2}\sum_{i=1}^2 g_{(i)}^2
 \left(\phi^{\dagger j}_A T^{a_i}\phi_{Aj}-{\phi}^j_BT^{a_i}
 {\phi}^\dagger_{Bj}\right)^2\nonumber\\
 &&\left.+\mbox{superpotential terms}
\right\}
 \nonumber\\
 s_\mu &=& -\sum_{i=1}^2\sigma_{\nu\rho}
 \gamma_\mu \lambda^{a_i} G^{(i)a_i\nu\rho}
+\frac{1}{2}\left(D^\nu \phi_A\right)^{\dagger j}\gamma_\nu
(1+\gamma_5)\gamma_\mu \psi_j
-\frac{1}{2}\widetilde{D}^\nu\phi_B^j  \gamma_\nu
(1-\gamma_5)\gamma_\mu \psi_j \nonumber\\
&& +\frac{1}{2} \gamma_\nu (1-\gamma_5)\gamma_\mu
\left(C\overline{\psi}^{Tj}\right) D_\nu\phi_{Aj} -\frac{1}{2}
\gamma_\nu (1-\gamma_5)\gamma_\mu
\left(C\overline{\psi}^{Tj}\right)\left(\widetilde{D}_\nu\phi_B
\right)^\dagger
 \nonumber\\
&& -\frac{1}{3}i\sigma_{\mu\nu}\partial^\nu \left[
\phi_{Aj}(1+\gamma_5)\left(C\overline{\psi}^{Tj}\right)
+\phi_A^{j\dagger} (1-\gamma_5)\psi_j\right]\nonumber\\
&& +\frac{1}{3}i\sigma_{\mu\nu}\partial^\nu \left[
\phi_{Bj}^\dagger(1-\gamma_5) \left(C\overline{\psi}^{Tj}\right)
+\phi_B^j (1+\gamma_5)\psi_j
  \right]
 \nonumber\\
 &&+2\sqrt{2}\sum_{i=1}^2 (-1)^{i+1}g_{(i)}\gamma_5\gamma_\mu
 \lambda^{a_i}\left[\phi_A^{j\dagger}T^{a_i}\phi_{Aj}
 -\phi_{Bj}T^a\phi_B^{\dagger j}  \right]
 + \mbox{superpotential terms};\nonumber\\
j_\mu &=&\frac{1}{2}\sum_{i=1}^2\overline{\lambda}^{a_i}
\gamma_\mu\gamma_5 \lambda^{a_i}-\frac{1}{3}\overline{\psi}^j
\gamma_\mu\gamma_5 \psi_j -\frac{2i}{3}\left[\phi_A^{\dagger j}
D_\mu \phi_{Aj}-
\left(D_\mu\phi_A\right)^{\dagger j} \phi_{Aj} \right]\nonumber\\
&&-\frac{2i}{3}\left[\phi_B^{j} \left(\widetilde{D}_\mu
\phi_{Bj}\right)^\dagger- \left(\widetilde{D}_\mu\phi_B\right)^j
\phi_{Aj}^\dagger \right]+\mbox{superpotential terms}.
\label{cus2}
\end{eqnarray}
In above equation, $\lambda^a$ and $\psi$ are Majorana and Dirac
spinors, respectively,
\begin{eqnarray}
\lambda^a=\left(\begin{array}{c}\lambda_\alpha^a\\
\overline{\lambda}^{a\dot{\alpha}} \end{array}\right),~~
\psi_j=\left(\begin{array}{c}\psi_{Aj\alpha}\\
\overline{\psi}^{\dot{\alpha}}_{Bj} \end{array}\right),
~~\left(C\overline{\psi}^T\right)^j
=\left(\begin{array}{c}\psi_{B\alpha}^j\\
\overline{\psi}^{j\dot{\alpha}}_A \end{array}\right).
\label{curr1}
\end{eqnarray}

Classically the above conservative currents satisfy
$\theta^\mu_{~\mu}=\gamma^\mu s_\mu=0$ up to the classical
superpotential terms. At quantum level  the superconformal anomaly
arises due to the non-vanishing $\beta$-functions of two gauge
couplings. We first analyze the chiral $U_R(1)$ symmetry anomaly.
Eq.\,(\ref{curr1}) shows that the chiral current $j^\mu$ is composed
of gluinoes $\lambda^{a_i}$, $\left(\psi_{Aj}\right)_{~r}^m$ and
$\left(\psi_B^j\right)_{~m}^{r}$. Relative to the first gauge group
$SU(N+M)$, there are $2N$ flavor matters (counting $m$ index in
$\left(\psi_{Aj}\right)_{~r}^m$ and
$\left(\psi_B^j\right)_{~m}^{r}$), which contribute $2N\times
(-1/2)=-N$ to the anomaly coefficient. The gluino $\lambda^{a_1}$ is
in the adjoint representation of $SU(N+M)$ and hence makes the
contribution $C_2[SU(N+M)]=N+M$ to the anomaly coefficient. So for
the first gauge field background, the chiral anomaly coefficient is
$N+M-N=M$. A similar analysis for the second gauge group $SU(N)$
gives the chiral anomaly coefficient  $-M$.  Therefore, we obtain
the chiral $U_R(1)$ anomaly,
\begin{eqnarray}
\partial_\mu j^\mu &=&\frac{M}{16\pi^2}\left(g_1^2 G_{\mu\nu}^{a_1}
\widetilde{G}^{a_1\mu\nu}-g_2^2 G_{\mu\nu}^{a_2}
\widetilde{G}^{a_2\mu\nu}\right) +\mbox{classical superpotential
part}. \label{chira2}
\end{eqnarray}

 To observe how the scale anomaly arises, we first consider the
 $SU(N)\times SU(N)$ gauge theory with chiral superfields
 $A_j$ in $(N,\overline{N})$ and $B^j$ in $(\overline{N},N)$
 representations. This theory is a superconformal field theory
in the sense that it has IR fixed-points. From the
 NSVZ $\beta$-functions for those two gauge couplings,
\begin{eqnarray}
\beta (g_1^2)&=&-\frac{g^3_1}{16\pi^2}\left[ 3N-2N (1-\gamma
(g))\right],\nonumber\\
\beta (g_2^2)&=& \frac{g^3_2}{16\pi^2}\left[ 3N-2 N(1-\gamma
(g))\right],
\end{eqnarray}
we can see that the zero-points of $\beta$-functions arise at the
anomalous dimension $\gamma(g)=-1/2$.  However, for the
$SU(N+M)\times SU(N)$ gauge theory with chiral superfields
 $A_j$ in $(N+M,\overline{N})$ and $B^j$ in $(N,\overline{N+M})$
 representations, the IR fixed points are removed since now
 the corresponding $\beta$-functions become
\begin{eqnarray}
\beta (g_1^2)&=&-\frac{3Mg^3_1}{16\pi^2},~~~~ \beta (g_2^2)=
\frac{3Mg^3_2}{16\pi^2}. \label{bfunction}
\end{eqnarray}
So the superconformal anomaly should arise near the IR fixed points
of the $M=0$ case . We have the scale anomaly similar to that listed
in Eq.\,(\ref{sanomaly1}),
\begin{eqnarray}
\theta^{\mu}_{~\mu}&=&\frac{3M}{8\pi^2} \sum_{i=1}^2
(-1)^{i+1}{g_{i}^2}\left[-\frac{1}{4}G^{(i)a_i\mu\nu}
G^{(i)a_i}_{\mu\nu}
+\frac{1}{2}i\overline{\lambda}^{a_i}\gamma^{\mu}
\left(\nabla_{\mu}{\lambda}\right)^{a_i} \right]\nonumber\\
&& +\mbox{classical superpotential part}. \label{scal2}
\end{eqnarray}
Further, using the beta functions (\ref{bfunction}) near the
IR-fixed points in the $M=0$ case, we easily find the $\gamma$-trace
anomaly of supersymmetry current,
\begin{eqnarray}
\gamma^\mu s_\mu &=& \frac{3M}{8\pi^2}
\sum_{i=1}^2(-1)^{i+1}g_i^2\left(-\sigma^{\mu\nu}G^{a_i}_{\mu\nu}
+\gamma_5 D^{a_i}\right)\lambda^{a_i}.
\end{eqnarray}

Let us turn to the manifestation of above superconformal anomaly in
the quantum effective action. We consider the local part of the
gauge invariant quantum effective action composed only of gauge
fields,
\begin{eqnarray}
{\Gamma}_{\rm eff}&=& \frac{1}{32\pi}\int d^2\theta\,\sum_{i=1}^2
\mbox{Im}\left[\Sigma_{(i)\,\rm eff} \mbox{Tr} \left(W^{(i)\alpha}
W^{(i)}_\alpha\right) \right]+\mbox{h.c.},
\end{eqnarray}
where $\Sigma_{(i)\,\rm eff}$ is given  in (\ref{supercou}). A
similar discussion as the case of ${\cal N}=1$ supersymmetric gauge
theory shows that the superconformal anomaly manifest itself  as the
 running of
gauge couplings $g_i$ and the shifts of both CP-angles
$\theta_i^{\rm CP}$ and the superpartner $\chi_{(i)}$  in the above
quantum effective action,
\begin{eqnarray}
\frac{1}{g^2_{(i)\,{\rm eff}}(q^2)}&=&\frac{1}{g^2_{(i)}(q_0^2)}
+(-1)^{i+1}\frac{3M}{8\pi^2}\ln
\frac{q_0}{q},  \nonumber\\
\theta^{\rm (CP)}_{(i)\,{\rm eff}} &=& \theta^{\rm
(CP)}_{(i)}+(-1)^{i+1}M, \nonumber\\
\chi_{(i)\,{\rm eff}} &=&
0+(-1)^{i+1}M.
 \label{effcou1}
\end{eqnarray}
This ends our discussion on the superconformal anomaly in the ${\cal
N}=1$ $SU(N+M)\times SU(N)$ supersymmetric gauge theory.

\section{Superconformal Anomaly from Brane Dynamics}
\label{sect3}

In this section we reveal the origin of superconformal anomaly in
the brane configuration. This will pave the way for us to
investigate the supergravity dual of superconformal anomaly of
supersymmetric gauge theory.

\subsection{Brane configuration for ${\cal N}=1$ supersymmetric
$SU(N+M)\times SU(N)$ gauge theory and K-S solution}
\label{subsect31}

The brane configuration for the ${\cal N}=1$ supersymmetric
$SU(N+M)\times SU(N)$ gauge theory has been constructed as the
following \cite{klwi}. One starts from the bulk  space-time composed
of the direct product of a four-dimensional Minkowski space $M^4$
and a six-dimensional conifold ${\cal C}_6$ whose base is
$T^{1,1}=\left[SU(2)\times SU(2)\right]/U(1)$. The isometry group of
$T^{1,1}$ is $SU(2)\times SU(2)\times U(1)$. Such a target
space-time  allows not only a stack of $N$ $D3$-branes moving freely
in the transverse space, but also a stack of $M$ fractional
$D3$-branes fixed at the singularity, the apex of ${\cal C}_6$. All
these $D3$-branes extend out in  $M^4$. Topologically $T^{1,1}\sim
S^2\times S^3$, and the fractional $D3$-branes can be considered as
$D5$-branes wrapped around two-cycle $S^2$.

The four-dimensional ${\cal N}=1$ supersymmetric gauge theory
produced from above brane configuration can be found by observing
the massless spectrum of open string with 4 Neumann and 6 Dirichlet
boundary conditions in $M^4\times {\cal C}_6$ \cite{klwi}. In the
case with no fractional $D3$ branes,  it turned out that the
resultant field theory on the world volume of $D3$-brane is the
${\cal N}=1$ $SU(N)\times SU(N)$ super-Yang-Mills theory coupled
with four chiral matter superfields, two of which
$\left(A_i\right)_{~r}^{m}$
 transform as $(N,\overline{N})$ and the other two $\left(B^j\right)_{~m}^r$
 as $(\overline{N},N)$ under the product gauge group
$SU(N)\times SU(N)$, $i=1,2$, $r=1,\cdots,N$,
$m=1,\cdots,\overline{N}$. The global symmetry group $SU_L(2)\times
SU_R(2)\times U_R(1)$ of the gauge theory comes from the isometry
group of $T^{1,1}$.
 When $M$ fractional branes are added at the apex of ${\cal C}_6$,
 the superconformal symmetry of this field model at the IR fixed
 points disappears and the resultant field theory is $SU(N+M)\times SU(N)$
 super-Yang-Mills theory coupled with chiral superfields
 $\left(A_i\right)_{~r}^m$ and $\left(B^j\right)_{~m}^r$ in
 the bifundmental representations
 $(N+M,\overline{N})$ and $(\overline{N+M},N)$ of the gauge group.

 On the  gravity side, when no fractional $D3$-branes are present,
 the three-brane solution
to type IIB supergravity corresponding to such a brane configuration
takes the following form \cite{klwi},
\begin{eqnarray}
ds^2 &=& H^{-1/2}(r)\eta_{\mu\nu}dx^\mu dx^\nu+H^{1/2}(r)\left
(dr^2+
r^2 ds^2_{T^{1,1}}\right),\nonumber\\
F_{(5)}&=&{\cal F}_{(5)}+\widetilde{\cal F}_{(5)}, \nonumber\\
 H(r) &=& 1+\frac{L^4}{r^4}, ~~ L^4=\frac{27\pi g_s
N(\alpha^\prime)^2}{4},\nonumber\\
ds^2_{T^{1,1}} &=&
\frac{1}{9}\left(2d\beta+\sum_{i=1}^2\cos\theta_i
d\phi_i\right)^2+\frac{1}{6}\sum_{i=1}^2\left(d\theta_i^2+\sin^2\theta_i
d\phi_i^2\right)\nonumber\\
&= &\frac{1}{9}
(\sigma^{\widetilde{3}}+\sigma^{\widehat{3}})^2+\frac{1}{6}
\sum_{{\widetilde{1}}=1}^2
(\sigma^{\widetilde{i}})^2+\frac{1}{6}\sum_{\widehat{i}=1}^2
(\sigma^{\widehat{i}})^2  = \frac{1}{9}
(g^5)^2+\frac{1}{6}\sum_{m=1}^4 (g^m)^2,
\nonumber\\
{\cal F}_{(5)} &=& \frac{1}{2}\pi\alpha^{\prime 2}N  \sin\theta_1
\sin\theta_2 d\beta\wedge d\theta_1\wedge d\theta_2\wedge
d\phi_1\wedge d\phi_2, \label{nfsolu1}
\end{eqnarray}
In above solution, $g^m$ are one-form bases on $T^{1,1}$,
\begin{eqnarray}
g^1 &\equiv&
\frac{\sigma^{\widetilde{1}}-\sigma^{\widehat{1}}}{\sqrt{2}},
~~~g^2\equiv\frac{\sigma^{\widetilde{2}}
-\sigma^{\widehat{2}}}{\sqrt{2}}, ~~~
g^3\equiv\frac{\sigma^{\widetilde{1}}
+\sigma^{\widehat{1}}}{\sqrt{2}}, \nonumber\\
g^4 &\equiv &
\frac{\sigma^{\widetilde{2}}+\sigma^{\widehat{2}}}{\sqrt{2}}, ~~~~
g^5\equiv\sigma^{\widetilde{3}}+\sigma^{\widehat{3}} ,
\end{eqnarray}
and
\begin{eqnarray}
\sigma^{\widetilde{1}}& {\equiv}&\sin\theta_1d\phi_1,
~~\sigma^{\widehat{1}}{\equiv}\cos 2\beta\sin\theta_2d\phi_2-\sin
2\beta d\theta_2,~~
 \sigma^{\widetilde{2}} {\equiv} d\theta_1,\nonumber\\
 \sigma^{\widehat{2}}&{\equiv}& \sin 2\beta\sin\theta_2d\phi_2
 +\cos 2\beta d\theta_2,~~
\sigma^{\widetilde{3}}\equiv \cos\theta_1d\phi_1, ~~
\sigma^{\widehat{3}}\equiv 2d\beta+\cos\theta_2 d\phi_2 .
\end{eqnarray}
All other fields in the solution vanish. Near the horizon limit
($r\rightarrow 0$), this solution yields $AdS_5\times T^{1,1}$
background geometry for type IIB superstring.

When the $D3$ branes are switched on, the resultant background  is
the K-S solution to type IIB supergravity \cite{klst}. This solution
is composed of not only the ten-dimensional space-time metric, the
self-dual five-form field strength $F_{(5)}$, but also the NS-NS-
and R-R two-forms, $B_{(2)}$ and $C_{(2)}$, living only on
$T^{1,1}\sim S^2\times S^3$ \cite{klst},
\begin{eqnarray}
ds_{10}^2 &=& h^{-1/2}(\tau) dx_{1,3}^2+ h^{1/2}(\tau) d s_6^2,\nonumber\\
\overline{F}_{(5)}&=& {\cal F}_{(5)}+\widetilde{F}_{(5)},~~~{\cal
F}_{(5)}=dx^0\wedge
dx^1\wedge dx^2\wedge dx^3\wedge dh^{-1},\nonumber\\
B_{(2)}&=&\frac{g_sM\alpha^\prime}{2}\left[ f(\tau)g^1\wedge
g^2+k(\tau)
g^3\wedge g^4\right], \nonumber\\
F_{(3)}&=&\frac{M\alpha^\prime}{2}\left\{g^3\wedge g^4\wedge
g^5+d\left[ F(\tau)\left(g^1\wedge g^3+g^2\wedge g^4\right) \right]
\right\} \label{ksso}
\end{eqnarray}
In above equation $h(\tau)$ is the warp factor, $ds_6^2$ is the line
element on the deformed conifold \cite{klst,rama},
\begin{eqnarray}
ds_6^2&=&\frac{1}{2}(12)^{1/3} K(\tau)\left\{\frac{1}{3[K(\tau)]^3}
\left[ d\tau^2+
(g^5)^2\right]\right.\nonumber\\
&&\left.+\sinh^2\left(\frac{\tau}{2}\right)\left[ (g^1)^2+(g^2)^2
\right]+\cosh^2\left(\frac{\tau}{2}\right)\left[ (g^3)^2+(g^4)^2
\right] \right\}, \nonumber\\
K(\tau) &=& \frac{(\sinh 2\tau-2\tau)^{1/3}}{2^{1/3}\sinh\tau}, ~~
F(\tau) = \frac{\sinh\tau-\tau}{2\sinh\tau},\nonumber\\
f(\tau) &=& \frac{\tau\coth\tau-1}{2\sinh\tau}(\cosh\tau-1),~~
k(\tau) =
\frac{\tau\coth\tau-1}{2\sinh\tau}(\cosh\tau+1),\nonumber\\
h(\tau) &=&C (g_sM)^2\frac{2^{2/3}}{4}\int_\tau^\infty dx
\frac{x\coth x-1}{\sinh^2x}\left[\sinh (2 x)-2x \right]^{1/3},
\label{kssoe}
\end{eqnarray}
$C$  being a normalization constant. This solution is non-singular
since the apex of the  conifold is resolved by the fractional brane
fluxes \cite{klst}.

At large $\tau$, the radial coordinate $\tau$ is related to the
radial coordinate $r$ as $r^3\sim 12^{1/2}e^\tau$. Consequently, the
solution reduces to the Klebanov-Tseytlin (K-T) solution
\cite{tseytlin},
\begin{eqnarray}
ds_{10}^2 &=& h^{-1/2}(r) dx_{1,3}^2+ h^{1/2}(r)
\left(dr^2+r^2 ds^2_{T^{1,1}}\right),\nonumber\\
{\cal F}_{(5)} &=& N_{\rm eff}(r) \omega_2 \wedge \omega_3, ~~
F_{(3)} = d C_{(2)}= c (r) \omega_3, ~~B_{(2)}=b(r) \omega_2, \nonumber\\
 h(r)&=&\frac{27\pi (\alpha^\prime)^2}{4r^4}\left[ g_s
N+\frac{3}{2\pi}(g_s
M)^2\ln\left(\frac{r}{r_0}\right)+\frac{3}{8\pi} (g_sM)^2\right],
\nonumber\\
\omega_2 &=&\frac{1}{2}\left(g^1\wedge g^2+g^3\wedge g^4\right) =
\frac{1}{2}\left(\sin\theta_1 d\theta_1\wedge d\phi_1-
\sin\theta_2 d\theta_2\wedge d\phi_2\right), \nonumber\\
\omega_3 &=& g^5\wedge \omega_2,~~
 \int_{S^2}\omega_2 = 4\pi, ~~\int_{S^3}\omega_3=8\pi^2,
 \nonumber\\
 b(r)&=& \frac{3g_s
 M\alpha^\prime}{2}\ln\left(\frac{r}{r_0}\right),~~
 c(r)=\frac{M\alpha^\prime}{2}, \nonumber\\
 N_{\rm eff}(r)&=& N+\frac{3}{2\pi} g_sM^2
 \ln\left(\frac{r}{r_0}\right),~~
 C_{(2)} = M\alpha^\prime \beta \omega_2~~\mbox{ (locally)}.
 \label{ksso3}
\end{eqnarray}
All other fields vanish. Obviously, this solution describes a
deformed $AdS_5\times T^{1,1}$ geometry. Eq.\,(\ref{ksso3}) shows
explicitly that there are $M$ units of R-R three-form fluxes passing
through the 3-cycle $S^3$ of $T^{1,1}$.

\subsection{Fractional Brane as the Origin for
Superconformal Anomaly} \label{subsect32}

We use the brane probe technique \cite{cliff} to show that presence
of fractional branes  in the brane configuration is the origin of
the superconformal anomaly origin.
 The low-energy dynamics of a stack of $Dp$-brane system
describes the dynamical behavior of open string modes trapped on the
world-volume of $Dp$-branes and their interaction with bulk
supergravity. For a single $D$-brane, the low-energy effective
action consists of the Dirac-Born-Infeld action (in Einstein
framework) \cite{cliff}
\begin{eqnarray}
S_{\rm DBI}&=&-\tau_{p} \int_{V_{p+1}} d^{p+1}x e^{(p-3)\phi/4}
\sqrt{-\det\left[G_{\mu\nu}+ e^{-\phi/2}
\left(B_{\mu\nu}+2\pi\alpha^\prime F_{\mu\nu}\right)\right]}.
 \label{dbiaction}
\end{eqnarray}
 and the Wess-Zumino (WZ) term,
\begin{eqnarray}
S_{\rm WZ}=\mu_{p}\int_{V_{p+1}} \left(C\wedge
e^{B+2\pi\alpha^\prime F}\right)_{p+1}. \label{wzterm}
\end{eqnarray}
The former describes the interaction between a $Dp$-brane and NS-NS
fields of type II superstring and the later is about the $Dp$-brane
interacting with R-R fields.
 The parameters
$\tau_p$ and $\mu_p$ are the tension and R-R charge of a $Dp$-brane,
which are actually equal because of the BPS-saturation feature of
$Dp$-brane,
\begin{eqnarray}
 \tau_p=\mu_p=\frac{1}{(2\pi)^p
{\alpha^{\prime}}^{(1+p)/2}},
\end{eqnarray}
 As for other
fields, $F_{\mu\nu}$ is the strength of $U(1)$ gauge field living on
the world-volume of $Dp$-branes;
 $G_{\mu\nu}$, $B_{\mu\nu}$ and
 $C_{\mu_1\mu_2\cdots\mu_n}$ are the pull-backs of ten-dimensional
bulk metric $G_{MN}$, the antisymmetric NS-NS field $B_{(2)MN}$ and
R-R fields  $C_{M_1\cdots M_n}$ to the $p+1$-dimensional
world-volume,
\begin{eqnarray}
&& G_{\mu\nu}=G_{MN}\partial_\mu X^{M} \partial_\nu X^{N}, ~~
B_{\mu\nu}=B_{MN}\partial_\mu X^{M} \partial_\nu X^{N},\nonumber\\
&& C_{\mu_1\mu_2\cdots\mu_n}=C_{M_1M_2\cdots M_n}\partial_{\mu_1}
X^{M_1}\partial_{\mu_2} X^{M_2}\cdots \partial_{\mu_n} X^{M_n}, ~~
n\leq p+1.
\end{eqnarray}

With the DBI action and WZ term, we now demonstrate how the
superconformal anomaly originates from the $D3$ fractional branes.
 First, when the transverse space like
conifold has singularity, the closed string states consist of both
the untwisted and twisted sectors \cite{dive}. The fractional
$Dp$-branes frozen at the singular point of transverse space are
identical to the $D(p+2)$-branes wrapped on two-cycle ${\cal C}_2$
like $S^2$ and the singular point can be considered as a vanishing
two-cycle. The fields corresponding to the twisted string states
emitted by fractional $Dp$-branes should locate at the singular
point of target space-time. Thus the twisted fields can always be
decomposed into two parts, one part living on the $p+1$-dimensional
world-volume of fractional $Dp$-brane, and the other on the blow-up
of the vanishing two-cycle ${\cal C}_2$ \cite{dive},
\begin{eqnarray}
V_{p+3}=V_{p+1}(x)\times {\cal C}_2, ~~~ B_{(2)}=B (x)\omega_2,
~~~C_{(p+3)}=C_{(p+1)}(x)\wedge \omega_2. \label{fracfield}
\end{eqnarray}
In above equation the twisted scalar field $B(x)$ and the pull-back
$C_{\mu_1\cdots\mu_{p+1}}$ of R-R antisymmetric tensor fields
 live on the
world-volume of $Dp$-branes.

Second, we employ the $D$-brane probe technique to observe the
dynamical behavior of the supersymmetric gauge theory implied from
the $D$-brane dynamics. The basic idea of brane probe technique is
the following \cite{cliff}. In order to detect the dynamics of a
supersymmetric $SU(N)$ gauge theory living on the world-volume of a
stack of $N$ coincident $Dp$-branes, one just places a probe brane
of the same type near those coincident branes. This brane
configuration will yield an $SU(N+1)$  supersymmetric gauge theory
with spontaneous breaking of gauge symmetry $SU(N+1)\rightarrow
SU(N)\times U(1)$. The justification for this is that the
coordinates of transverse space are identified as scalar fields on
$Dp$-brane world-volume. So when we move one brane some distance
from the coincident $Dp$-branes, this means  that the scalar fields
get vacuum expectation values. If we consider the Coulomb branch of
the supersymmetric $SU(N+1)$ gauge theory, the probe brane decouples
from those coincident $Dp$-branes after symmetry breaking. However,
according to the decoupling theorem, at the energy scale
characterized by the distance of the probe brane away from those
coincident branes, the coupling of
 supersymmetric $U(1)$  gauge theory on the world-volume
of probe brane should equal to the gauge coupling of $SU(N)$ gauge
theory on the world-volume of those $N$ coincident D-branes.
Therefore, the probe brane action can provide us the information
about the running of gauge coupling and  shift of the $\theta$-angle
of the supersymmetric
 $SU(N)$ gauge theory living on the world-volume of coincident $Dp$-branes.

On the other hand, since a stack of $N$ coincident $Dp$-branes
modify the background space-time, the probe technique can be
considered as a probe brane moving slowly
 in the supergravity background produced by the coincident branes
 to be probed. Therefore, we  can use the DBI action and WZ term
 for the probe brane in the $p$-brane solution background produced by the
 $N$ coincident $Dp$-branes to observe the dynamical behavior of
 a $p+1$-dimensional $SU(N)$
 supersymmetric gauge theory. In
 this way one can easily get the quantum information of a supersymmetric
  gauge theory defined on the world-volume of those coincident $Dp$-branes.

Based on the brane probe technique, we substitute the fields in
(\ref{fracfield}) into the DBI action (\ref{dbiaction}) and the WZ
term (\ref{wzterm}) for a single brane and expand them to the
quadratic terms of the gauge field strength,
\begin{eqnarray}
&& S_{\rm DBI}+S_{\rm WZ}\nonumber\\
&=&-\tau_{p+2} \int_{V_{p+1}\times {\cal C}_2} d^{p+3}x
e^{(p-1)\phi/4} \sqrt{-\det\left[G_{\mu\nu}+ e^{-\phi/2}
\left(B_{ab} +2\pi\alpha^\prime F_{\mu\nu}\right)\right]}
\nonumber\\
&&  + \mu_{p+2}\int_{V_{p+1}\times {\cal C}_2} \left[C\wedge\left(
e^{B+2\pi\alpha^\prime
F}\right)\right]_{(p+1)+2}\nonumber\\
&=&-\tau_{p+2} \int_{V_{p+1}} d^{p+1}x\, e^{(p-3)\phi/4}
\sqrt{-\det\left(G_{\mu\nu}+ 2\pi\alpha^\prime e^{-\phi/2}
F_{\mu\nu}\right)}\int_{{\cal C}_2} B_{(2)}\nonumber\\
&&+\mu_{p+2}\left[\int_{V_{p+1}} C_{(p+1)}\int_{{\cal C}_{2}}\,
C_{(2)}+\frac{1}{2}(2\pi\alpha^\prime)^2 \int_{V_{p+1}}
C_{(p-3)}\wedge\left(F\wedge F\right) \int_{{\cal
C}_2}\,C_{(2)}\right.\nonumber\\
&&+\left.\int_{V_{p+1}} C_{(p+1)}\int_{{\cal C}_2}\,
B_{(2)}+\frac{1}{2}(2\pi\alpha^\prime)^2 \int_{V_{p+1}}
C_{(p-3)}\wedge\left(F\wedge F\right) \int_{{\cal C}_2}\,B_{(2)}
+\cdots\right]
\nonumber\\
&=& S_{\rm brane-bulk}+ S_{\rm gauge}; \nonumber\\
&& S_{\rm gauge} =-\alpha^\prime \tau_p \int_{V_{p+1}} d^{p+1}x\,
e^{(p-3)\phi/4} \sqrt{-\det \left( G_{\mu\nu}\right)}
\left[-\frac{1}{4}\left(F^{\lambda\rho}
F_{\lambda\rho}\right)\right]\,e^{-\phi} \int_{{\cal C}_2} B_{(2)} \nonumber\\
&& + \frac{1}{2}\alpha^\prime \mu_p \left[\int_{V_{p+1}} \left(
F\wedge F\right)\wedge C_{(p-3)}\,\left(\int_{{\cal C}_2} C_{(2)}
+\int_{{\cal C}_2} B_{(2)}\right) \right]+\cdots,
\nonumber\\
 && \mu, \nu =0,\cdots, p; ~~~\alpha,\beta=0,\cdots,p, a,b;
 \label{expdbi}
\end{eqnarray}
where $a,b=1,2$  are the indices on the vanishing 2-cycle ${\cal
C}_2$. Note that in Eq.\,(\ref{expdbi})
 we choose $X^I$=Constant ($I=p+1,\cdots,9$),
 and hence $G_{IJ}=0$. Recalling the relation between the gauge couplings and
string couplings,
\begin{eqnarray}
\frac{4i\pi}{g^2_{\rm YM}}+\frac{\theta^{\rm (CP)}}{2\pi}=i
e^{-\phi}+\frac{C_{(0)}}{2\pi},
\end{eqnarray}
and  choosing $p=3$, we can  see  from Eq.\,(\ref{expdbi}) that it
is the NS-NS- and R-R two-form fluxes $\int_{{\cal C}_2} B_{(2)}$,
$\int_{{\cal C}_2} C_{(2)}$ carried by fractional branes through the
shrunken 2-cycle that have created the running of $U(1)$ gauge
coupling and the shift of $\theta$-angle. This exactly leads to
 the superconformal anomaly of the supersymmetric gauge theory on
 a stack of coincident $D$-branes. Therefore,
we conclude that in a brane configuration the fractional branes
frozen at the singular point of background space-time is the origin
of superconformal anomaly of a supersymmetric gauge theory.

\section{Manifestation of Superconformal Anomaly on K-S Solution}
 \label{sect4}

\subsection{Breaking of $U(1)$ rotational
and scale symmetries  in transverse space by fractional $D3$-branes}
\label{subsect41}

In the following we focus on the brane configuration with $N$ bulk
$D3$-branes and $M$ fractional $D3$-branes in target space-time
$M^4\times {\cal C}_6$ and analyze how the fractional brane affects
the space-time symmetry of $AdS_5\times T^{1,1}$.

The classical scale symmetry and chiral $U_R(1)$-symmetry of the
$SU(N+M)\times SU(M)$
 supersymmetric
gauge theory living on the world-volume of $D3$-branes comes from
the geometrical scale- and rotation transformation  invariance  of
the transverse space,
\begin{eqnarray}
x^I\longrightarrow \mu e^{i\alpha}x^I, ~~4\leq I\leq 10.
\end{eqnarray}
Now let us observe  how the above scale- and $ SO(2)\cong U(1)$
rotation symmetries are reflected in the K-S solution. First, in the
case with no fractional branes, the near-horizon ($r\to 0$) limit of
the three-brane solution (\ref{nfsolu1}) is $AdS_5\times T^{1,1}$,
\begin{eqnarray}
 ds_{10}^2 &=& \frac{r^2}{L^2}\eta_{\mu\nu} dx^\mu  dx^\nu
+\frac{L^2}{r^2} dr^2+L^2 ds^2_{T^{1,1}}\nonumber\\
ds^2_{T^{1,1}} &=& \frac{1}{9}\left(2d\beta+\sum_{i=1}^2\cos\theta_i
d\phi_i\right)^2+\frac{1}{6}\sum_{i=1}^2\left(d\theta_i^2+\sin^2\theta_i
d\phi_i^2\right)\nonumber\\
 F_{(5)} &=& {\cal F}_{(5)}+{}^\star{\cal F}_{(5)}
=\frac{1}{2}\pi\alpha^{\prime 2}N \left[\omega_2\wedge \omega_3
+{}^\star\left(\omega_2\wedge \omega_3\right)\right]\nonumber\\
&=& \frac{r^3}{g_s L^4} dx^0 \wedge dx^1 \wedge dx^2 \wedge dx^3
\wedge dr + \frac{L^4}{27 g_s} g^1\wedge g^2\wedge g^3\wedge
g^4\wedge g^5,\nonumber\\
 L^2 &=& \frac{3\sqrt{3\pi g_sN}\alpha^\prime}{2}.
 \label{at11}
\end{eqnarray}
This solution shows explicitly that the scale symmetry is trivially
present in the transverse space part and the $U(1)$ symmetry is
reflected in the invariance  under  the $\beta$-angle rotation,
$\beta\longrightarrow \beta+\alpha$. Then we switch on fractional
branes, the solution
 (\ref{ksso3}) shows that the metric ceases to be
$AdS_5\times T^{1,1}$. The reason for this is that $h(r)$ and
$F_{(5)}$ get
 logarithmic  dependence on the radial coordinate,
\begin{eqnarray}
  h(r)&\sim& \frac{1}{r^4} \left[C_1+C_2\ln
 \frac{r}{r_0}\right], \nonumber\\
  \overline{F}_5 &\sim& \left[N+C_3\ln \frac{r}{r_0}\right] [
 \omega_2\wedge\omega_3+{}^\star(\omega_2\wedge\omega_3)],
 \end{eqnarray}
 where the coefficients $C_i$  ($i=1,2,3$) can be extracted out from the solution
(\ref{ksso3}). Further,
 the background fields $B_{(2)MN}$ and $C_{(2)MN}$
 are generated by the fractional brane fluxes.

  In Ref.\,\cite{kow}, it was analyzed that the non-invariance
 of $C_{(2)}$ under the $\beta$-angle rotation leads to
the chiral R-symmetry anomaly in ${\cal N}=1$ supersymmetric
$SU(N+M)\times SU(N)$ gauge theory. We make a similar analysis to
show that  the non-invariance of $B_{(2)}$ under scale
transformation in transverse space  results in the scale anomaly.
The $B_{(2)}$ in Eq.\,(\ref{ksso}) under the scale transformation
$r\rightarrow \mu r$ transforms as
\begin{eqnarray}
B_{(2)}\longrightarrow
B_{(2)}+\frac{3g_sM\alpha^\prime}{2}\omega_2\ln \mu . \label{bfv}
\end{eqnarray}
According to the relation between string coupling and gauge
couplings of the ${\cal N}=1$ supersymmetric $SU(N+M)\times SU(N)$
gauge theory \cite{klst,klwi},
\begin{eqnarray}
\frac{1}{g_1^2}+\frac{1}{g_2^2}\sim e^{-\phi}\sim \frac{1}{g_s},
~~~\frac{1}{g_1^2}-\frac{1}{g_2^2}\sim e^{-\phi} \left[
\int_{S^2}B_{(2)}-\frac{1}{2}\right],
\end{eqnarray}
one has
\begin{eqnarray}
\frac{1}{g_1^2}\sim \frac{1}{2g_s}\left[
\int_{S^2}B_{(2)}-\frac{1}{2}\right], ~~~\frac{1}{g_1^2}\sim
\frac{1}{2g_s}\left[-\int_{S^2}B_{(2)}+\frac{3}{2}\right].
\label{twogc}
\end{eqnarray}
A comparison between (\ref{bfv}) and (\ref{twogc}) shows that the
variation of $B_{(2)}$ under scale transformation is relevant to
$\beta$-functions for the
 $ SU(N+M)\times SU(N)$ gauge couplings at the IR fixed points of the
 ${\cal N}=1$ $SU(N)\times SU(N)$ supersymmetric gauge theory \cite{klst},
\begin{eqnarray}
 \frac{d}{d(\ln \mu)}\frac{8\pi^2}{g_1^2(\mu)}\sim 3M, ~~~
\frac{d}{d(\ln \mu)}\frac{8\pi^2}{g_2^2(\mu)}\sim -3M.
\end{eqnarray}
This exactly lead to the  scale anomaly coefficients shown in
Eq.\,(\ref{scal2}).

\subsection{Supersymmetry Breaking
due to Modification  on $AdS_5\times T^{1,1}$ by Fractional Branes}
\label{subsect42}

In this subsection  we use the result of  Ref.\,\cite{rama} to
emphasize how one-half of supersymmetries break in type IIB
supergravity due to the modification on space-time background  by
fractional branes. The standard method of investigating
supersymmetry breaking produced by fractional branes is to check the
Killing spinor equations obtained from  supersymmetry
transformations for the fermionic fields of type IIB supergravity in
the K-S solution background (\ref{ksso}) and count the number of
Killing spinors. The bosonic field content of type IIB supergravity
 consists of the dilaton field $\phi$, the metric
$G_{MN}$ and  the second-rank antisymmetric tensor field $B_{(2)MN}$
in the NS-NS sector, the axion field $C_{(0)}$, the two-form
potential $C_{(2)MN}$ and  four-form potential $C_{(4)MNPQ}$ with
self-dual field strength in the R-R sector. The fermionic fields are
left-handed complex Weyl gravitino $\Psi_M$ and right-handed complex
Weyl dilatino ${\Lambda}$, $\Gamma^{11}\Psi_M=-\Psi_M$,
$\Gamma^{11}{\Lambda}={\Lambda}$. The supersymmetry transformations
for the fermionic fields read \cite{schw},
\begin{eqnarray}
\delta \Lambda &=& \frac{i}{\kappa_{\rm 10}}\Gamma^M
\epsilon^{\star} P_M -\frac{i}{24}\Gamma^{MNP}\epsilon
G_{(3)MNP}+\mbox{fermions~relevant
~terms}, \nonumber\\
\delta \Psi_M &=& \frac{1}{\kappa_{\rm
10}}\left(D_M-\frac{1}{2}iQ_M\right)\epsilon+\frac{i}{480}
\Gamma^{PQRST}\Gamma_M \epsilon \widehat{F}_{(5)PQRST} \nonumber\\
&& +\frac{1}{96} \left(\Gamma_M^{~~NPQ}G_{(3)NPQ}
-9\Gamma^{NP}G_{(3)MNP}\right)
\epsilon^{\star}+\mbox{fermions relevant terms},\nonumber\\
&& M,N,\cdots =0,1,\cdots, 9. \label{susytr1}
\end{eqnarray}
In above equation, the  supersymmetry transformation parameter
$\epsilon$ is a left-handed complex Weyl spinor,
$\Gamma^{11}\epsilon=-\epsilon$, and other quantities are listed as
the following \cite{schw,poch},
\begin{eqnarray}
P_M &=& f^2\partial_M B, ~~ Q_M=f^2\mbox{Im}\left( B\partial_M
B^{\star}\right),~~ B\equiv \frac{1+i\tau}{1-i\tau}, \nonumber\\
 f^{-2} &=& 1-B B^{\star},~~\tau=C_{(0)}+ie^{-\phi},\nonumber\\
 G_{(3)MNP}&=& f\left(\widehat{F}_{(3)MNP}
 -B \widehat{F}_{(3)MNP}^\star\right), \nonumber\\
  \widehat{F}_{(3)MNP} &=&
3\partial_{[M}A_{(2)NP]},~~
A_{(2)MN}\equiv C_{(2)MN}+iB_{(2)MN},\nonumber\\
\widehat{F}_{(5)MNPQR}&=& F_{(5)MNPQR}-\frac{1}{8}\times 10\,
\kappa_{\rm 10}\mbox{Im} \left(A_{[(2)MN} \widehat{F}_{(3)PQR]}
\right)\nonumber\\
&=&{F}_{(5)MNPQR}-\frac{5}{4}\kappa_{\rm 10} C_{[(2)MN}{H}_{(3)PQR]}
+\frac{5}{4} \kappa_{\rm 10} B_{[(2)MN}{F}_{(3)PQR]} ,
\nonumber\\
F_{(5)MNPQR} &=& 5\partial_{[M} C_{(4)NPQR]},\nonumber\\
 D_M\epsilon &=& \partial_M \epsilon +\frac{1}{4}\omega_M^{~~AB}\Gamma_{AB}
 \epsilon.
\end{eqnarray}
To highlight the effects of fractional branes, we first consider the
supersymmetry transformation in the background without fractional
branes, i.e., the background (\ref{nfsolu1}) furnished by
three-brane solution whose near-horizon limit is $AdS_5\times
T^{1,1}$. The supersymmetry transformations for the fermionic fields
in the $AdS_5\times T^{1,1}$ background read
\begin{eqnarray}
\delta \Lambda &=&0, \nonumber\\
\delta \Psi_M &=&\frac{1}{\kappa_{\rm
10}}\left(\partial_M+\frac{1}{4}
\omega_M^{~~AB}\Gamma^{AB}\right)\epsilon
+\frac{i}{480}\Gamma^{PQRST} {F}_{(5)PQRST}\Gamma_M\epsilon=0.
\label{susytr2}
\end{eqnarray}
The first equation $\delta \Lambda=0$  is trivially satisfied in the
background (\ref{at11}). The Killing spinor equation $\delta \Psi_M
=0$ yields \cite{rama}
\begin{eqnarray}
\epsilon &=& r^{\Gamma_\star/2}\left[  1+\frac{\Gamma_r}{2R^2} x^\mu
\Gamma_\mu \left( 1-\Gamma_\star\right)\right]\epsilon_0,\nonumber\\
\Gamma_{\star} &\equiv&
i\Gamma_{x_0}\Gamma_{x_1}\Gamma_{x_2}\Gamma_{x_3},~~~ \Gamma_\star^2
=1, \label{killing1}
\end{eqnarray}
where $\epsilon_0$ is an arbitrary constant spinor in ten dimensions
but constrained by
\begin{eqnarray}
\Gamma_{g_1g_2}\epsilon_0=\epsilon_0,
~~~\Gamma_{g_3g_4}\epsilon_0=-\epsilon_0. \label{cons1}
\end{eqnarray}
In Eqs.\,(\ref{killing1}) and (\ref{cons1}) the coordinates of
$AdS_5\times T^{1,1}$ are used
 to denote the components of $\Gamma$-matrices for clarity.
The constraint (\ref{cons1}) on $\epsilon_0$ leads to
$\Gamma_{g_1g_2}\epsilon=\epsilon$,
$\Gamma_{g_3g_4}\epsilon=-\epsilon$. Therefore, the Killing spinor
$\epsilon$ has eight independent components.

The Killing spinor (\ref{killing1}) is actually a unified expression
for the following two types of Killing spinors \cite{rama,lpt},
\begin{eqnarray}
\epsilon_+&=& r^{1/2}\epsilon_{0+},
~~~\epsilon_-=r^{-1/2}\epsilon_{0-}+\frac{r^{1/2}}{L^2}\Gamma_r
x^\mu
\Gamma_{\mu}\epsilon_{0-},\nonumber\\
\epsilon_{0\pm} &=&\frac{1}{2}\left(1\pm
\Gamma_\star\right)\epsilon_0,
~~~\Gamma_\star\epsilon_{0\pm}=\pm\epsilon_{0\pm}.
 \label{killing2}
\end{eqnarray}
Eq.\,(\ref{killing2}) shows that $\epsilon_+$ is independent of the
coordinates on $D3$-brane world-volume  and is a right-handed
eigenspinor of $\Gamma_\star$, i.e.,
$\Gamma_\star\epsilon_+=\epsilon_+$. It thus represents ${\cal N}=1$
Poinc\'{a}re supersymmetry in the dual supersymmetric gauge theory.
Further, $\epsilon_-$ depends on $x^\mu$ linearly and is not an
eigenspinor of $\Gamma_\star$. So it characterizes ${\cal N}=1$
conformal supersymmetry of the dual supersymmetric gauge theory in
four dimensions.

 When the $M$ fractional branes are switched on, the
 space-time background is described by the K-S solution (\ref{ksso}).
 The Killing spinor equations for dilatino and gravitino become
\begin{eqnarray}
\delta {\Lambda}
&=&-\frac{i}{24}\widehat{F}_{(3)MNP}\Gamma^{MNP}\epsilon
=0, \nonumber\\
\delta \Psi_M &=&\frac{1}{\kappa}\left(\partial_M+\frac{1}{4}\,
\omega_M^{~~AB}\Gamma^{AB}\right)\eta+\frac{i}{480}
\Gamma^{PQRST}\Gamma_M \epsilon \widetilde{F}_{(5)PQRST} \nonumber\\
& +&\frac{1}{96} \left(\Gamma_M^{~~NPQ}\widehat{F}_{(3)NPQ}
-9\Gamma^{NP}\widehat{F}_{(3)MNP}\right) \epsilon^{\star}
 = 0. \label{susytr3}
\end{eqnarray}
 The same procedure with the metric (\ref{ksso}) as in the case
 with no fractional brane
 leads to the following Killing spinor \cite{rama}
\begin{eqnarray}
\epsilon=h^{-1/8}(\tau) \exp\left(-\frac{\alpha}{2}\Gamma_{g_1g_2}
\right)\epsilon_0,
\end{eqnarray}
where $\alpha$ is determined by $\sin\alpha= -{1}/{\cosh\tau}$,
$cos\alpha = {\sinh\tau}/{\cosh\tau}$. $\epsilon_0$ is a constant
spinor constrained by \cite{rama},
\begin{eqnarray}
\Gamma_\star\epsilon_0=-i\epsilon_0,
~~~\Gamma_{g_1g_2}\epsilon_0=-\Gamma_{g_3g_4}\epsilon_0, ~~\Gamma_{r
g_5} \epsilon_0=-i\epsilon_0. \label{kicon1}
\end{eqnarray}
Hence the Killing spinor $\epsilon$ satisfies
\begin{eqnarray}
\Gamma_\star \epsilon=-i\epsilon,
~~~\Gamma_{g_1g_2}\epsilon=-\Gamma_{g_3g_4}\epsilon=
i\left(\cos\alpha+\sin\alpha\Gamma_{g_1g_3}\right)\epsilon,~~\Gamma_{rg_5}
\epsilon=-i\epsilon. \label{kicon2}
\end{eqnarray}
These three constraints determines that $\epsilon$ has only four
independent components. In particular, $\epsilon$ is independent of
the coordinate on the world-volume of $D3$-branes. Therefore, the
four-component Killing spinor $\epsilon$ means the only existence of
${\cal N}=1$ Poin\'{a}re supersymmetry in the dual supersymmetric
gauge theory. This indicates that the conformal supersymmetry in the
dual four-dimensional supersymmetric gauge theory collapses due to
the presence of fractional $D3$-branes.

\section{Superconformal anomaly as spontaneously breaking of
local  supersymmetry in gauged $AdS_5$ supergravity and super-Higgs
Mechanism} \label{sect5}

\subsection{Generality}
\label{subsect51}

The discussions  in last section have shown that in comparison with
$AdS_5\times T^{1,1}$ case, the K-S solution background fails to
preserve some of local symmetries in type IIB supergravity. We
choose the K-S solution as a classical vacuum configuration for type
IIB supergravity and observe the theory around such a background. In
the case without fractional branes, the near-horizon limit of the
three-brane solution
 is $AdS_5\times T^{1,1}$ (cf.\,(\ref{at11})).
However, a  gravitational system has geometrical meaning, expanding
ten-dimensional type IIB supergravity around $AdS_5\times T^{1,1}$
is actually a process of performing spontaneous compactification of
type IIB supergravity \cite{freu,kim,mark} on $T^{1,1}$. The
resultant theory after compactification should be ${\cal N}=2$
five-dimensional $U(1)$ gauged $AdS_5$ supergravity coupled with
${\cal N}=2$ $SU(2)\times SU(2)$ Yang-Mills vector multiplets and
several Betti tensor supermultiplets, whose origin is due to the
nontrivial topology of $T^{1,1}$ \cite{ita}. The local symmetries in
${\cal N}=2$ $U(1)$ gauged $AdS_5$ supergravity consist of ${\cal
N}=2$ supersymmetry, $SO(2,4)$ isometry symmetry and $U(1)$ gauge
symmetry.

When fractional branes switch on,  the space-time background becomes
the K-S solution and
 its UV limit, the K-T solution (\ref{ksso3}) is a
 deformed $AdS_5\times T^{1,1}$. Consequently, the
 isometry symmetry of
the deformed $AdS_5\times T^{1,1}$ and  the supersymmetry it
preserves is less than that exploited from $AdS_5\times T^{1,1}$.
Therefore, the compactified theory obtained from the
compactification of type IIB supergravity on  the deformed
$T^{1,1}$, i.e., a certain five-dimensional gauged supergravity,
should possess less local symmetries than those extracted from
$AdS_5\times T^{1,1}$. This means that some of local symmetries
break spontaneously in the gauged $AdS_5$ supergravity since the
symmetry loss originates from the vacuum configuration. Further, as
is well known, a physical consequence of spontaneous breaking of
local symmetry is the occurrence of the Higgs mechanism. To show
this phenomenon clearly, just like what usually done in gauge
theory, we reparametrize the field variable and ``shift" the vacuum
configuration described by the K-T solution back to $AdS_5\times
T^{1,1}$. The essence of this operation is performing local symmetry
transformation and transferring the non-symmetric feature of vacuum
configuration to the classical action. Then when we expand type IIB
supergravity around $AdS_5\times T^{1,1}$ with newly defined field
variables, the action of the five-dimensional gauged supergravity
 should lose some of local symmetries and the graviton multiplet
in gauged $AdS_5$ supergravity should acquire a mass by eating a
Goldstone multiplet relevant to  NS-NS- and R-R two-form fields in
the K-S solution. In this way, we reveal how the super-Higgs
mechanism due to the spontaneous breaking of local supersymmetry in
the gauged $AdS_5$ supergravity occurs.
 This phenomenon was actually somehow noticed in the Kaluza-Klein
supergravity \cite{duff}: when the internal manifold is deformed or
squashed, it keeps less symmetries  for the compactified theory than
the undeformed  internal manifold. This is the so-called ``space
invader" scenario and
 can be naturally given an interpretation in terms of
 spontaneous breaking of local symmetry \cite{duff}.

Based on above analysis, we first consider the case without
fractional brane and  observe type IIB supergravity in $AdS_5\times
T^{1,1}$ background. This will lead to ${\cal N}=2$ $U(1)$ gauged
$AdS_5$ supergravity coupled to $SU(2)\times SU(2)$ Yang-Mills
vector multiplets and Betti scalar-, vector- and tensor multiplets
\cite{ita} in which the graviton supermultiple is massless. Then we
 go to the case with fractional branes and see how
 the graviton multiplet  becomes massive  due to the
 spontaneous breaking of local supersymmetry in  gauged
 $AdS_5$ supergravity.

\subsection{Type IIB supergravity in
$AdS_5 \times T^{1,1}$ Background and Gauged $AdS_5$ Supergravity}
\label{subsect52}

We first give a brief review on the compactification of type IIB
supergravity on $T^{1,1}$ \cite{ita}. In general, when performing
compactification of a certain $D$-dimensional supergravity on
$AdS_{D-d}\times K^d$, one should first linearize the classical
equation of motion of the field function $\Phi (x,y)\equiv
\Phi^{\{J\}[\lambda]}(x,y)$ in $AdS_{D-d}\times K^d$ background. In
above, $K^d=G/H$ is certain $d$-dimensional compact Einstein
manifold, $x$ and $y$ are the coordinates on $AdS_{D-d}$ and $K^d$,
respectively; $\{J\}$ and $[\lambda]$ denote the representations of
local Lorenz groups $SO(2,D-d-1)$ and $SO(d)$ realized on the field
function. The next step is to expand  $y$-dependent part of the
field function in terms of $H$-harmonics on $K^d$, which are
representations of the group $G$ branched with respect to its
subgroup $H$. If the internal manifold is $K^d=S^d=SO(d+1)/SO(d)$,
the maximally symmetric Einstein manifold, the expansion procedure
works straightforwardly since $H$ is $SO(d)$, which is precisely the
local Lorenz group of the $d$-dimensional internal manifold.
However, if the internal manifold is a less symmetric one, and $H$
is  a subgroup of $SO(d)$. Then the field function representations
of $SO(d)$ are usually reducible with respect to $H$. Therefore,
only those $H$-harmonics that are identical to the $SO(d)$ field
function representations branched by $H$ can contribute to the
expansion of field functions on $K^d$.

The compactification of type IIB supergravity on $T^{1,1}$ is
exactly this case. $T^{1,1}$ is the coset space $G/H=[SU(2)\times
SU(2)]/U_{\rm H}(1)$ and  the generator of $U_{\rm H}(1)$ is the sum
$T^3+\widehat{T}^3$ of two diagonal generators $T^3$ and
$\widehat{T}^3$ of $SU(2)\times SU(2)$. The harmonics on $T^{1,1}$
are representations of $SU(2)\times SU(2)$ labeled  by the weight
$\{\nu\}=(j,l)$,
\begin{eqnarray}
Y(y)\equiv \left(\left[Y^{(j,l,r)} (y)\right]^m\right).
\end{eqnarray}
where $m=1,\cdots, (2j+1)\times (2l+1)$, which labels the
representation of $SU(2)\times SU(2)$, and $r$ is the  quantum
number for the $U(1)$ group whose generator is $T_3-\widehat{T}_3$.
The representations $(j,l)$ are reducible with respect to the
subgroup $U_{\rm H}(1)$ and hence decompose into a direct sum of
fragments graded by the $U_{\rm H}(1)$-charge $q_i$,
\begin{eqnarray}
\left[Y^{(j,l,r)} (y)\right]^m=\bigoplus_i\,\left[Y^{(j,l,r)}
(y)\right]^m_{~q_i}=\left(\begin{array}{c}
\left[Y^{(j,l,r)}(y)\right]^m_{~q_1}\\ \left[Y^{(j,l,r)}
(y)\right]^m_{~q_2}\\ \vdots\\
\left[Y^{(j,l,r)} (y)\right]^m_{~q_N}
\end{array} \right).
\end{eqnarray}
 The irreducible representations
$\left[Y^{(j,l,r)} (y)\right]^m_{~q_i}$ are called $U_{\rm
H}(1)$-harmonics on $T^{1,1}$.

 On the other
hand, the field function $\Phi^{\{J\}[\lambda]}(x,y)$  on
$AdS_5\times T^{1,1}$   belongs to a certain representation of the
local Lorentz group $SO(2,4)\times SO(5)$, $\{J\}$, $[\lambda]$
denoting representation wights for $SO(2,4)$ and $SO(5)$,
respectively. The $U_{\rm H}(1)$ is a subgroup of $SO(5)$ because
the representation of its generator is naturally embedded into the
representation of $SO(5)$ generators \cite{sast}. The $SO(5)$
representations $[\lambda]$  furnished by
$\left[X^{[\lambda]}(y)\right]^n$, the $y$-dependent part of
$\Phi^{\{J\}[\lambda]} (x,y)$, are also reducible with respect to
$U_{\rm H}(1)$, where $n$ labels the representation of $SO(5)$ in
field function space. The field function representation space thus
decomposes into a direct sum of irreducible subspaces labeled by the
$U_{\rm H}(1)$-charge $q_\xi$,
\begin{eqnarray}
 \left[X^{[\lambda]}(y)\right]^n
=\bigoplus_{\xi=1}^K\left[X^{[\lambda]}(y)\right]^n_{~q_\xi}
=\left(\begin{array}{c} \left[X^{[\lambda]}(y)\right]^n_{~q_1}\\
\left[X^{[\lambda]}
(y)\right]^n_{~q_2}\\ \vdots\\
\left[X^{[\lambda]} (y)\right]^n_{~q_K}
\end{array} \right).
\end{eqnarray}
 The irreducible representations
$\left[X^{[\lambda]}(y)\right]^n_{~q_\xi}$ in above equation are
called the $SO(5)$ harmonics. The field function  admits a natural
expansion in terms of $SO(5)$ harmonics.
 Therefore,
 an $U_{\rm H}$-harmonics $Y$ can contribute to the expansion of a
field function on $T^{1,1}$ only when it
 is identical to (or contains)  certain $SO(5)$ harmonics $X$.
  A detailed analysis on how the $SU(2)\times SU(2)$ representations
branched with respect to $U_{\rm H}(1)$ appear in the decomposition
 of the $SO(5)$ field function representations
 with respect to $U_{\rm H}(1)$ is shown in
Ref.\,\cite{ita}.

 Once  the expansion of  field functions
$\Phi^{\{J\}[\lambda]} (x,y)$ in terms of those admissible $U_{\rm
H}(1)$-harmonics on $ T^{1,1}$  is known. That is  \cite{ita},
\begin{eqnarray}
\Phi^{\{J\}[\lambda]} (x,y) &=&\left(\Phi^{\{J\}}_{ab\cdots}
(x,y)\right)^{[\lambda]} =\bigoplus_{i=1}^N
\left(\Phi^{\{J\}}_{q_i}\right)^{[\lambda]}= \left(\begin{array}{c}
\Phi^{\{J\}}_{q_1}(x,y)\\  \Phi^{\{J\}}_{q_2}(x,y)\\ \vdots\\
\Phi^{\{J\}}_{q_N}(x,y)
\end{array} \right),
 \nonumber\\
 \Phi^{\{J\}}_{q_i}(x,y) &=& \sum_{j,l}\sum_m
\sum_r \Phi_{q_i m}^{\{J\}(j,l,r)}(x) \left[Y^{(j,l,r)}
(y)\right]^m_{~q_i}, \label{harmexp1}
\end{eqnarray}
one substitutes it into the  linearized
 equations of motion of type IIB
supergravity in  $AdS_5 \times T^{1,1}$ background,
\begin{eqnarray}
\left( K_x^{\{J\}}+K_y^{[\lambda]}\right)\Phi^{\{J\}[\lambda]}
(x,y)=0. \label{lineareq}
\end{eqnarray}
In Eqs.\,(\ref{harmexp1}) and (\ref{lineareq}), $a,b$ are the
indices for the $[\lambda]$-representation, $K_x^{\{J\}}$ and
$K_y^{[\lambda]}$ are the kinetic operators of type IIB supergravity
on $AdS_5$ and $T^{1,1}$, respectively.  There are three types
kinetic operators $K_y^{[\lambda]}$ in supergravity: the Hodge-de
Rahm- and
 Laplacian operators acting on scalar-, vector- and antisymmetric
fields, the Dirac and Rarita-Schwinger operator acting on the
fermionic fields and the Lichnerowicz operator  on the symmetric
rank-two graviton field. All these three types of operators  are (or
are related to) certain Laplace-Beltrami operators on $T^{1,1}$.
Thus the action of
 $K_y^{[\lambda]}$ on  $U_{\rm H}(1)$-harmonics gives \cite{ita}
 \begin{eqnarray}
K_y^{[\lambda]}\left[Y^{(j,l,r)}
(y)\right]^m_{~q_i}=M_{ik}^{(j,l,r)}\left[Y^{(j,l,r)}
(y)\right]^m_{~q_k}.
\end{eqnarray}
This equation together with the linearized equation of motion
(\ref{lineareq}) and $U_{\rm H}(1)$-harmonic expansion
(\ref{harmexp1}) of field functions determines that
$M_{ij}^{(j,l,r)}$ are mass matrices for the K-K particle tower
$\Phi_{q_k m}^{\{J\}(j,l,r)}(x)$ in $AdS_5$ space,
 \begin{eqnarray}
\left(\delta_{ik}K_x^{\{J\}}+ M_{ik}^{(j,l,r)}\right)\Phi_{q_k
m}^{\{J\}(j,l,r)}(x)=0.
\end{eqnarray}
Therefore, the zero modes of the kinetic operator $K_y^{[\lambda]}$
 constitute  the field content of ${\cal N}=2$ $U(1)$ gauged $AdS_5$
supergravity coupled with $SU(2)\times SU(2)$ Yang-Mills vector
supermultiplets as well as some  Betti tensor multiplets \cite{ita}.

We then turn to the explicit form of $AdS_5\times T^{1,1}$. As a
standard step in Kaluza-Klein theory,  to perform the
compactification on $T^{1,1}$, one should write the metric
(\ref{nfsolu1}) in the Kaluza-Klein metric form,
\begin{eqnarray}
ds^2&=& h^{-1/2}(r) \eta_{\mu\nu}dx^\mu dx^\nu+h^{1/2})(r)dr^2\nonumber\\
&+& h^{1/2}(r)
r^2\left\{\frac{1}{9}\left(g^5-2A\right)^2+\frac{1}{6}\left[\sum_{r=1}^{2}
\left(g^r-K^{ar}W^a\right)^2+\sum_{s=3}^4\left(g^s-L^{bs}
\widetilde{W}^b \right)^2\right]\right\}. \label{kkmetric}
\end{eqnarray}
In Eq.\,(\ref{kkmetric}), $K^{ar}$ and $L^{bs}$ are the components
of Killing vectors $K^a$ and $L^b$ on $T^{1,1}$, and they convert
into the generators of $SU(2)\times SU(2)$ gauge group in the
resultant ${\cal N}=2$ $U(1)$ gauged supergravity coupled with six
Yang-Mills vector supermultiplets,
\begin{eqnarray}
[K^{a_1},K^{a_2}]=i\epsilon^{a_1a_2a_3}K^{a_3},
~~~[L^{b_1},L^{b_2}]=i\epsilon^{b_1b_2b_3}L^{b_3};
\end{eqnarray}
$W^a=W^{a}_{~\alpha} dx^\alpha$ and
$\widetilde{W}^{b}=\widetilde{W}^{b}_{~\alpha}dx^\alpha$ are the
corresponding six $SU(2)$ gauge fields; $A=A_\alpha dx^\alpha$ is
the $U(1)$ gauge field corresponding to the isometric symmetry
$U(1)$, and it constitutes an ${\cal N}=2$ supermultiplets with the
graviton $h_{\alpha\beta}$ and gravitini $\psi_\alpha^i$, $i=1,2$,
in the gauged $AdS_5$ supergravity.

As a straightforward consequence of using the K-K metric
(\ref{kkmetric}), the self-dual five-form  in Eq.\,(\ref{nfsolu1})
should be modified to keep its self-duality with respect to the K-K
metric \cite{kow},
\begin{eqnarray}
\overline{F}_5 &=& d\overline{C}_4=\frac{1}{g_s}
\partial_r h^{-1}(r) dx^0\wedge dx^1\wedge dx^2\wedge dx^3
\wedge dr\nonumber\\
&&+\frac{L^4}{27}\left[\chi\wedge g^1 \wedge g^2 \wedge g^3 \wedge
g^4-dA\wedge g^5\wedge dg^5+\frac{3}{L}\left({}^{\star_5}
dA\right)\wedge dg^5 \right]. \label{kkfiveform}
\end{eqnarray}
Locally, there exists
\begin{eqnarray}
 \overline{C}_4 &=& \frac{1}{g_s}
h^{-1}(r) dx^0\wedge dx^1\wedge
dx^2\wedge dx^3\nonumber\\
&+&\frac{2L^4}{27}\left[\beta g^1 \wedge g^2 \wedge g^3 \wedge
g^4-\frac{1}{2}A\wedge g^5\wedge
dg^5+\frac{3}{2r}h^{-1/4}(r)\left({}^{\star_5} dA\right)\wedge
dg^5\right], \label{kkfiveform2}
\end{eqnarray}
where $\chi=g^5-2A$, ${\star_5}$ is the
 five-dimensional Hodge dual operator defined with respect to
 $AdS_5$ metric
\begin{eqnarray}
ds_{{\rm AdS}_5}^2=g_{\alpha\beta}^{(0)}dx^\alpha
dx^\beta=\frac{r^2}{L^2} \eta_{\mu\nu}dx^\mu dx^\nu+
\frac{L^2}{r^2}dr^2, ~~~\alpha,\beta=0,1,\cdots,4. \label{adsme}
\end{eqnarray}
  Obviously, both the K-K metric (\ref{kkmetric}) and
$\overline{F}_5$ are gauge invariant under local $U(1)$ gauge
transformation in gauged $AdS_5$ supergravity,
\begin{eqnarray}
\beta\rightarrow \beta+\varphi, ~~~  A\rightarrow A+d\varphi.
\end{eqnarray}
$\overline{F}_5$ satisfies the Bianchi identity $d\overline{F}_5=0$,
which determines $A$ is a massless vector fields in $AdS_5$ space,
i.e., $d{}^{\star_5}dA=0$.

Finally, we briefly mention how the graviton supermultiplet in the
gauged $AdS_5$ supergravity comes from the compactification of type
IIB supergravity on $T^{1,1}$ \cite{ita}. First,
$h_{\alpha\beta}(x)$ is the zero-mode in the scalar $U_{\rm
H}(1)$-harmonic expansion of $h_{\alpha\beta}(x,y)$, which is the
$AdS_5$ space-time component
 of ten-dimensional graviton $h_{MN}(x,y)$;
 $A_{\alpha}$ is the linear combination of two zero-modes in the vector
$U_{\rm H}(1)$-harmonic expansion of $h_{\alpha a}$ and
$C_{(4)\alpha abc}$, which are the crossing  components on $AdS_5$
and $T^{1,1}$ of the graviton $h_{MN}(x,y)$ and the R-R four form
potential $C_{(4)MNPQ}(x,y)$, respectively; $\psi_\alpha$ comes from
the massless mode in the spinor $U_{\rm H}(1)$-harmonics expansion
of $\psi_\alpha$, which is the $AdS_5$  component of ten-dimensional
gravitino $\psi_M$ \cite{ita}.  The quadratic  action  for the
graviton supermultiplet of ${\cal N}=2$ $U(1)$ gauged $AdS_5$
supergravity  can be straightforwardly extracted out from the
compactification of type IIB supergravity on $AdS_5\times T^{1,1}$,
\begin{eqnarray}
 S=\frac{1}{\kappa^2_5} \int d^5x \sqrt{-g} \left(-\frac{1}{2} R-\frac{1}{2}
\overline{\psi}^i_\alpha\gamma^{\alpha\beta\gamma}
D_\beta\psi_{i\gamma}
-\frac{3L^2}{32}F^{\alpha\beta}F_{\alpha\beta}-\frac{6}{L^2}+\cdots
\right). \label{fiveaction}
\end{eqnarray}
 It shows that the graviton multiplet
$(A_\alpha, \psi_\alpha^i,h_{\alpha\beta})$ is massless.
%According
%to the holographic definition on
%AdS/CFT correspondence \cite{witt1}, this graviton multiplet  is
%dual to the current supermultiplet $(j_\mu , s_\mu, T_{\mu\nu})$ in
%a conformal invariant ${\cal N}=1$ supersymmetric gauge theory on
%the $AdS_5$ boundary.

 Now we come to the stage of revealing the dual of  superconformal
 anomaly in the ${\cal N}=1$ $SU(N+M)\times SU(N)$ supersymmetric gauge theory
with two flavors in the bifundmental representation of gauge group.

\subsection{Dual of Chiral $U_R(1)$-symmetry anomaly $\partial_\mu j^\mu$}
\label{subsect53}

In the following we  review briefly  how the dual of chiral $U_R(1)$
anomaly of
 ${\cal N}=1$ $SU(N+M)\times SU(N)$ gauge theory was found  in Ref.\,\cite{kow}
  from
type IIB supergravity in the K-S solution background.

When there is no fractional brane, the K-K metric and the five-form
field shown in Eqs.\,(\ref{kkmetric}) and (\ref{kkfiveform}) are
$U(1)$ gauge invariant  after the compactification on $T^{1,1}$ is
performed. The graviton multiplet of gauged $AdS_5$ supergravity is
massless. When fractional branes are present, both the K-K metric
(\ref{kkmetric}) and the five-form potential (\ref{kkfiveform}) get
deformed, but they still keep invariant under $U(1)$ gauge
transformation. However, Eq.\,(\ref{ksso}) shows that the field
strength $F_{(3)}$ of R-R two-form $C_{(2)}$
 arises in the background solution. Since $F_{(3)}$ depends linearly
on the angle $\beta$,
\begin{eqnarray}
 F_{(3)}=\frac{M\alpha^\prime}{2}\omega_3=\frac{M\alpha^\prime}{2}
g^5\wedge \omega_2=\frac{M\alpha^\prime}{4} g^5\wedge
\left(g^1\wedge g^2+g^3\wedge g^4\right),
\end{eqnarray}
it is not invariant under the rotation of $\beta$. As what usually
done in dealing with the spontaneous  breaking of gauge symmetry, we
shift $F_{(3)}$ as
\begin{eqnarray} \overline{F}_{(3)}
=\frac{M\alpha^\prime}{2} \left(g^5+2\partial_\alpha\theta
dx^\alpha\right) \wedge \omega_2.
\end{eqnarray}
by introducing a new field,
\begin{eqnarray}
 \theta\equiv F \int _{S^2} C_2,
 \label{fgold}
 \end{eqnarray}
where  $F$ is a field function in ten dimensions. The shifted
$\overline{F}_{(3)}$ is obviously invariant under the gauge
transformation
\begin{eqnarray}
 \beta\rightarrow \beta+\varphi, ~~~\theta\rightarrow
\theta-\varphi.
\end{eqnarray}
Further,  $\overline{F}_{(3)}$ can be re-expressed in terms of a
gauge invariant $\chi \equiv g^5-2A$ and  a newly defined  vector
field $W_\alpha$,
\begin{eqnarray}
\overline{F}_3 &=& \frac{M\alpha^\prime}{2} \left(\chi+2 W_\alpha
dx^\alpha\right) \wedge \omega_2
=\overline{F}_3^{(0)}+M\alpha^\prime W_\alpha dx^\alpha
\wedge \omega_2; \nonumber\\
W_\alpha & \equiv & A_\alpha+\partial_\alpha\theta,
~~~\overline{F}_3^{(0)}\equiv \frac{M\alpha^\prime}{2}\wedge
\omega_2.
 \label{rrthree}
\end{eqnarray}
The  vector field $W_\alpha$ and the $U(1)$ gauge field $A_\alpha$
have the same field strength $F_{\alpha\beta}$, but  $W_\alpha$ has
got a longitudinal component. We then consider type IIB supergravity
in the symmetric vacuum configuration furnished by the K-K metric
(\ref{kkmetric}), the self-dual five-form (\ref{kkfiveform}) and R-R
three-form field strength  $\overline{F}_3^{(0)}$ defined in
(\ref{rrthree}). The Einstein-Hilbert- and the R-R three-form terms
 in the classical action of type IIB supergravity,
 gives \cite{kow}
\begin{eqnarray}
S_{\rm IIB}&=&\frac{1}{2\kappa_{10}^2}\int d^{10}x \left[E
R-\frac{1}{2}{F}_3\wedge {}^\star
{F}_3+\cdots\right]\nonumber\\
&=&\frac{1}{2\kappa_{10}^2}\int d^{10}x E\left[-\frac{1}{9}
h^{1/2}(r) r^2 F^{\alpha\beta}F_{\alpha\beta}- \left(
\frac{3M\alpha^\prime}{h^{1/2}(r)r^2}\right)^2 W^\alpha W_\alpha
+\cdots\right].
\end{eqnarray}
This ten-dimensional action implies  that the $U(1)$ gauge field in
gauged $AdS_5$ supergravity,  which comes from the compactification
of type IIB supergravity on $T^{1,1}$, obtain a mass proportional to
the fluxes carried by fractional branes. This phenomenon is exactly
the Higgs effect in the following sense: when fractional branes are
present, the classical vacuum configuration is $U(1)$ symmetric.
After the classical solution is modified to make it gauge invariant,
the classical action of gauged $AdS_5$ supergravity around the
symmetric vacuum configuration should lose gauge symmetry and the
$U(1)$ gauge field acquires a mass.

\subsection{ Dual of  Scale Anomaly $\theta^{\mu}_{~\mu}$}
\label{subsect54}

In this subsection we investigate the gravity dual of scale anomaly
in a similar way as looking for the dual of chiral anomaly. It
should
 also
 correspond to a certain Higgs effect in gauged $AdS_5$ supergravity since it
shares a supermultiplet with the chiral $R$-symmetry anomaly on the
field theory side. We first analyze on gravity side which local
symmetry corresponding to the scale symmetry on field theory side
becomes broken in K-S solution. Obviously, this local symmetry is
not directly related to the isometry symmetry of internal manifold
$T^{1,1}$. It should lie in the diffeomorphism symmetry of $AdS_5$
space. The argument for this statement is implied from
Ref.\,\cite{imbi}, where it was shown that near the $AdS_5$ boundary
the diffeomorphism symmetry of $AdS_5$ metric decomposes into a
combination of four-dimensional diffeomorphism symmetry and Weyl
symmetry. The AdS/CFT correspondence at supergravity level tells us
that the four-dimensional diffeomorphism- and Weyl symmetries are
equivalent to the conservation and tracelessness of energy-momentum
tensor in the dual four-dimensional supersymmetric gauge theory
\cite{chch1}.
 This suggests that we should observe how the diffeomorphism
symmetry of  $AdS_5$ space is spoiled by  fractional branes to look
for the dual of scale anomaly.

As discussing chiral anomaly, we first observe the case without
fractional branes. Since we care about how the graviton is affected
by the breaking of diffeormorphism symmetry of $AdS_5$ space, so
only the Einstein-Hilbert- and the self-dual five-form terms  in
type IIB supergravity are selected out,
\begin{eqnarray}
S_{\rm IIB} =\frac{1}{2\kappa_{\rm 10}^2}\int d^{10}x E
\left(R-\frac{1}{4\times 5!}{F}_{(5)MNPQR}{F}^{~~MNPQR}_{(5)}
 \right)+\cdots.
\end{eqnarray}
Making the expansion
\begin{eqnarray}
G_{MN}&=&G^{(0)}_{MN}+h_{MN}, \nonumber\\
G^{MN}&=&G^{(0)MN}-h^{MN}+h^{MP}h_P^{~~N}+{\cal O}(h^3), \nonumber\\
\sqrt{-G} &=&\sqrt{-G_0}\left[
1+\frac{1}{2}h+\frac{1}{4}\left(\frac{1}{2} h^2-h^{MN} h_{MN}
\right)\right]+{\cal O}(h^3),\nonumber\\
R^{M}_{~~NPQ}
&=&R^{(0)M}_{~~~~NPQ}+\frac{1}{2}\left[\nabla_P,\nabla_Q \right]
h^M_{~~N}+\frac{1}{2}\left(\nabla_P \nabla_N h^M_{~~Q}-\nabla_Q
\nabla_N h^M_{~~P} \right)\nonumber\\
&& -\frac{1}{2}\left(\nabla_P \nabla^M h_{NQ}-\nabla_Q \nabla^M
h_{NP} \right)-\frac{1}{2}h^{MR}\left[\nabla_P,\nabla_Q \right]
h_{RN}\nonumber\\
&& -\frac{1}{2}h^{MR}\left(\nabla_P \nabla_N h_{RQ}-\nabla_Q
\nabla_N h_{RP} \right) +\frac{1}{2}h^{MR}\left(\nabla_P \nabla_R
h_{NQ}-\nabla_Q \nabla_R h_{NP} \right)\nonumber\\
&& +\frac{1}{4}\left(\nabla^R h^M_{~~P}-\nabla_P h^{MR}-\nabla^M
h^R_{~~P} \right)\left(\nabla_Q h_{RN}+\nabla_N h_{RQ}-\nabla_R
h_{NQ} \right)\nonumber\\
&&-\frac{1}{4}\left(\nabla^R h^M_{~~Q}-\nabla_Q h^{MR}-\nabla^M
h^R_{~~Q} \right)\left(\nabla_P h_{RN}+\nabla_N h_{RP}-\nabla_R
h_{NP} \right)\nonumber\\
&&+{\cal O}(h^3),
\end{eqnarray}
we obtain the quadratic action for the graviton of type IIB
supergravity in the $AdS_5\times T^{1,1}$ background,
\begin{eqnarray}
S_{\rm Q.G.}&=&\frac{1}{2\kappa_{10}^2}\int d^{10}x
\sqrt{-G^{(0)}}\left[-\frac{1}{4} \nabla^P h^{MN}  \nabla_P
h_{MN}+\frac{1}{2}\nabla^P h^{MN} \nabla_M h_{PN}\right.\nonumber\\
&&+\frac{1}{4}\nabla^M h\left( \nabla_M h-\frac{1}{2}
\nabla^N h_{MN}\right)\nonumber\\
&& + \left( R^{(0)}-\frac{1}{4\times
5!}F_{(5)}^{(0)MNPQR}F_{(5)MNPQR}^{(0)}
\right)\frac{1}{4}\left( \frac{1}{2}h^2-h^{ST}h_{ST} \right)\nonumber\\
 &&-\frac{1}{4\times 5!}\left(10 h^{M^\prime [M}h^{NN^\prime}
 F_{(5)MNPQR}^{(0)}F_{(5)M^\prime
N^\prime}^{(0)~~~~~PQR]}\right.\nonumber\\
&& \left.\left.-\frac{5}{2}h\,h^{M^\prime[M}
 F_{(5)MNPQR}^{(0)}F_{(5)M^\prime
}^{(0)~~~NPQR]} +5 h^{M^\prime
M^{\prime\prime}}h_{M^{\prime\prime}}^{~~~[M}
F_{(5)MNPQR}^{(0)}F_{(5)M^\prime}^{(0)~~NPQR}\right) \right]
\nonumber\\
&=&\frac{1}{2k_{10}^2}\int d^{10}x \sqrt{-G^{(0)}}\left[-\frac{1}{4}
\nabla^P h^{MN}  \nabla_P
h_{MN}+\frac{1}{2}\nabla^P h^{MN} \nabla_M h_{PN}\right.\nonumber\\
&&+\frac{1}{4}\nabla^M h\left( \nabla_M h-\frac{1}{2} \nabla^N
h_{MN}\right)
+\frac{1}{2}R_{MPNQ}h^{MN}h^{PQ}-\frac{1}{2}h^{MN}h_{N}^{~P} R_{PM}\nonumber\\
&&\left. +\frac{27}{32}\frac{\pi \alpha^{\prime 2}N}{h^2(r) r^5}
\left(h^{MN}h_{MN}- \frac{1}{2}h^2 \right) \right]\nonumber\\
&=&\frac{1}{2k_{10}^2}\int d^{10}x \sqrt{-G^{(0)}}\left[-\frac{1}{4}
\nabla^P h^{MN}  \nabla_P
h_{MN}+\frac{1}{2} \nabla_M h^{MN}  \nabla^P h_{PN}\right.\nonumber\\
&&\left.+\frac{1}{4}\nabla^M h\left( \nabla_M h-\frac{1}{2} \nabla^N
h_{MN}\right)-\frac{9}{32}\frac{\pi \alpha^{\prime 2}N}{h^2(r) r^5}
\left(h^{MN}h_{MN}+ \frac{1}{2}h^2 \right) \right].
\label{qudraction}
\end{eqnarray}
 The use of following identity is
made in deriving above quadratic action,
\begin{eqnarray}
\left[\nabla_M,\nabla_P\right]
h_N^{~P}=R_{MPN}^{~~~~~~Q}h_Q^{~P}-R_{MP}h_N^{~~P}.
\end{eqnarray}

 We then perform  the compactification of the quadratic action (\ref{qudraction})
 on $T^{1,1}$ .
 With gauge-fixing conditions $
D^\xi h_{\xi\alpha}=D^\xi h_{(\xi\varepsilon)}=0$, the expansion of
$h_{MN}(x,y)$ in terms of the $U_{\rm H}(1)$-harmonics on $T^{1,1}$
is the following \cite{ita},
\begin{eqnarray}
h_{MN}(x,y) & = &\left( h_{\alpha\beta} (x,y), h_{\alpha \xi}(x,y),
h_{\xi\varepsilon}(x,y) \right),\nonumber\\
h_{\alpha\beta} (x,y) &=& \sum_{j,l,r} h_{\alpha\beta}^{(j,l,r)} (x)
Y^{(j,l,r)}_0(y),
\nonumber\\
h_{\alpha \xi} (x,y) &=& \sum_{j,l,r} A^{(j,l,r)}_{\alpha} (x)
Y^{(j,l,r)}_\xi (y)=\sum_{i=1}^5\sum_{j,l,r_i}
A^{(j,l,r_i)}_{\alpha} (x) Y^{(j,l,r_i)}_{q_i} (y) ,
\nonumber\\
h_{(\xi\varepsilon)} (x,y) &\equiv&
h_{\xi\varepsilon}-\frac{1}{5}g_{\xi\varepsilon}h^\tau_{~\tau}=
\sum_{j,l,r} \varphi^{(j,l,r)} (x) Y^{(j,l,r)}_{(\xi\varepsilon)}(y)\nonumber\\
&=& \sum_{i=1}^{10}\sum_{j,l,r_i}\varphi^{(j,l,r_i)}
(x)Y^{(j,l,r_i)}_{q_i} (y),\nonumber\\
h^\xi_{~\xi} (x,y) &=& \sum_{j,l,r} \pi^{(j,l,r)} (x) Y^{(j,l,r)}_0
(y),
\end{eqnarray}
where $D_\xi$ is the $SO(5)$ covariant derivative on $T^{1,1}$.
 The compactification  on $T^{1,1}$ formally leads to
 the quadratic action for the graviton
 in  ${\cal N}=2$ $U(1)$ gauged $AdS_5$ supergravity,
\begin{eqnarray}
S_{\rm q.g.} &=&\frac{1}{2k_{5}^2}\int d^{5}x
\sqrt{-g^{(0)}}\left[-\frac{1}{4} \nabla^\gamma h^{\alpha\beta}
\nabla_\gamma
h_{\alpha\beta}+\frac{1}{2}\nabla^\gamma h^{\alpha\beta} \nabla_\alpha
h_{\gamma\beta}\right.\nonumber\\
&&\left.+\frac{1}{4}\nabla^\alpha h\left( \nabla_\alpha
h-\frac{1}{2} \nabla^\beta h_{\alpha\beta}\right)-\frac{1}{L^2}
\left(h^{\alpha\beta}h_{\alpha\beta}+ \frac{1}{2}h^2 \right)+\cdots
\right]. \label{afac}
\end{eqnarray}
The graviton $h_{\alpha\beta}(x)$  is the fluctuation around the
metric of $AdS_5$ space (\ref{adsme}),
\begin{eqnarray}
ds^2_5=g_{\alpha\beta} dx^\alpha
dx^\beta=\left[g_{\alpha\beta}^{(0)}+h_{\alpha\beta}(x)\right]
dx^\alpha dx^\beta. \label{conback}
\end{eqnarray}
The $SO(2,4)$ diffeomorphism invariance means the following
infinitesimal gauge symmetry for the graviton in $AdS_5$ space,
\begin{eqnarray}
\delta h_{\alpha\beta}=\nabla^{(0)}_\alpha\xi_\beta
+\nabla^{(0)}_\beta\xi_\alpha, \label{diff}
\end{eqnarray}
where the covariant derivative $\nabla^{(0)}_\alpha$ is defined with
respect to the metric $g^{(0)}_{\mu\nu}$ of $AdS_5$ space.  Near the
$AdS_5$ boundary, $\xi_\alpha =(\xi_\mu,\xi_r)$, the above bulk
diffeomorphism transformation (\ref{diff}) decomposes into
four-dimensional diffeormorphism- and  Weyl transformations,
$\xi_\mu$ and $\xi_r$ playing the roles of transformation
parameters, respectively \cite{imbi}. The action (\ref{afac})
determines the equation of motion for the graviton in $AdS_5$ space
\cite{higu},
\begin{eqnarray}
E_{\alpha\beta} &\equiv & \frac{1}{2}\left( \nabla^\gamma
\nabla_\gamma h_{\alpha\beta}-\nabla_\alpha\nabla_\gamma
h^\gamma_{~\beta}-\nabla_\beta\nabla_\gamma h^\lambda_{~\alpha}+
\nabla_\alpha\nabla_\beta h\right)\nonumber\\
&&+\frac{1}{2}g^{(0)}_{\alpha\beta}\left(\nabla_\gamma\nabla_\delta
h^{\gamma\delta}- \nabla^\gamma \nabla_\gamma h\right)-\frac{2}{L^2}
\left( h_{\alpha\beta}+\frac{1}{2}g^{(0)}_{\alpha\beta} h \right)
=0. \label{eomgr}
\end{eqnarray}
Due to the gauge symmetry (\ref{diff}), $h_{\alpha\beta}$ contain
non-physical modes and the physical ones should be the traceless and
transverse part of $h_{\alpha\beta}$. Eq.\,(\ref{eomgr}) leads to
the identity, $\nabla^{(0)}_\beta E^{\alpha\beta}=0$. This identity
and the gauge-fixing  condition $\nabla_\alpha^{(0)}
h^{\alpha\beta}=0$ fixes the  physical degrees of freedom of the
$AdS_5$ graviton.

 When  fractional $D3$-branes switch on, the $AdS_5\times T^{1,1}$
 background get deformed by R-R- and NS-NS three-form
 fluxes carried by fractional branes. The isometric
  symmetry of the deformed $AdS_5\times T^{1,1}$ reduces.
   So the compactified theory obtained from the action
   (\ref{qudraction}) in the deformed $AdS_5\times T^{1,1}$
   background
  should suffer from the spontaneous symmetry
  breaking from $SO(2,4)$ to $SO(1,4)$ and present the consequent Higgs effect.
In the following we demonstrate  how this Higgs mechanism occurs.

 The action of type IIB
supergravity composed of the   self-dual five-form $F_{(5)}$, the
NS-NS two-form $B_{(2)}$ and the Einstein-Hilbert part is
\begin{eqnarray}
 S_{\rm IIB}
&=&\frac{1}{2\kappa_{\rm 10}^2}\int d^{10}x\sqrt{- \overline{G}}
\left(\overline{R}-\frac{1}{2\times 3!} H_{(3)MNP} H^{MNP}_{(3)}
\right. \nonumber\\
&&\left.-\frac{1}{4\times 5!}{F}_{(5)MNPQR}{F}^{MNPQR}_{(5)}
 \right). \label{t2bs}
\end{eqnarray}
Note that for simplicity of discussion we do not consider terms
relevant to the R-R three-form. The Einstein equation from this
classical action is
\begin{eqnarray}
\overline{R}_{MN}&=&
\frac{1}{96}F_{(5)MPQRS}F_{(5)N}^{~~~~PQRS}\nonumber\\
&&+\frac{1}{4}\left( H_{(3)MPQ} H_{(3)N}^{~~~~PQ}-\frac{1}{12}G_{MN}
H_{(3)PQR} H^{(3)PQR} \right).
\end{eqnarray}
This equation  shows  that without three-form field strength
$H_{(3)PQR}$, the self-dual five-form flux leads to the three-brane
solution (\ref{nfsolu1}) whose near-horizon limit is $AdS_5\times
T^{1,1}$ if the flat limit is $M^4\times {\cal C}_6$ \cite{balo},
while the presence of NS-NS three-form flux makes a deformation on
$ASdS_5\times T^{1,1}$ and leads to a less symmetric vacuum
background. According to the basic idea of the Higgs mechanism, we
should make the vacuum configuration symmetric by performing a gauge
transformation which re-parametrises the field variable. This can be
done by shifting the Ricci curvature and absorbing the three-form
flux contribution into it,
\begin{eqnarray}
R\equiv \overline{R}-\frac{1}{2\times 3!} H_{(3)MNP}
H^{~~MNP}_{(3)}.
\end{eqnarray}
Consequently, the above Einstein equation becomes  the one with only
five-form fluxes as source,
\begin{eqnarray}
R_{MN}&\equiv & \overline{R}_{MN}-\frac{1}{4}\left( H_{(3)MPQ}
H_{(3)N}^{~~~~PQ}-\frac{1}{12}G_{MN} H_{(3)PQR} H^{PQR}_{(3)}
\right)\nonumber\\
& = &\frac{1}{96}F_{(5)MPQRS}F_{(5)N}^{~~~~PQRS}. \label{ree}
\end{eqnarray}
It gives the three-brane  solution (\ref{nfsolu1}), the symmetric
vacuum background. Eq.\,(\ref{ree}) implies that there exists
following connection between deformed and undeformed Riemannian
curvatures,
\begin{eqnarray}
R_{KMLN}&=&\overline{R}_{KMLN}-\frac{1}{4}\left[ H_{(3)KMQ}
H_{(3)LN}^{~~~~~~Q}\right.\nonumber\\
 &&\left.-\frac{1}{12\times
9}\left(G_{KL}G_{MN}-G_{KN}G_{ML}\right)H_{(3)PQR}
H^{PQR}_{(3)}\right].
\end{eqnarray}

We expand the gravitational field around the symmetric $AdS_5 \times
T^{1,1}$ vacuum configuration. However, just like the Higgs
phenomenon in gauge theory, $h_{MN}$ must undergo a gauge
transformation, which can  make $h_{MN}$ pick up a longitudinal
component and  become massive. This is, the following operation
should be made on background metrics and graviton fields,
\begin{eqnarray}
G_{MN}=\overline{G}_{MN}^{(0)}+{h}_{MN}={G}_{MN}^{(0)}+\overline{h}_{MN},
\label{reme}
\end{eqnarray}
where ${G}_{MN}^{(0)}$ and $\overline{G}_{MN}^{(0)}$ denote the
symmetric and deformed $AdS_5\times T^{1,1}$ metrics, respectively,
and $\overline{h}_{MN}$ and ${h}_{MN}$ are  graviton fluctuations
around these two space-time backgrounds, respectively.

The above discussion is a qualitative analysis in ten dimensions.
 In
the following we  use  the K-T solution to find the explicit shift
and make the Higgs phenomenon more clear.
 According to the K-T solution (\ref{ksso3}),
\begin{eqnarray}
ds_{\rm d-AdS_5}^2&=&\overline{g}_{\alpha\beta}^{(0)} dx^\alpha
dx^\beta  \approx  \frac{r^2}{L^2}\left[1-A(r)\right]\eta_{\mu\nu}
dx^\mu
dx^\nu +\frac{L^2}{r^2}\left[1+A(r)\right] dr^2, \nonumber\\
A(r)&=&\frac{3}{4\pi}\frac{M^2}{N}g_s\left(\frac{1}{4}+ \ln
\frac{r}{r_0}\right), \label{comf}
\end{eqnarray}
This deformed $AdS_5$ background shows that it is the logarithmic
dependence on the radial coordinate that breaks the Weyl symmetry
corresponding to scale symmetry on field theory side. Consequently,
the $SO(2,4)$ isometry symmetry of $AdS_5$ space breaks to
$SO(1,4)$.

We restore the deformed $AdS_5$ vacuum background back to $AdS_5$.
The $(\alpha\beta)$ component of Eqs.\,(\ref{ree})
 and the  explicit form of NS-NS two-form
$B_{(2)}$  given in (\ref{ksso3}) determine that there should exist
\begin{eqnarray}
\overline{g}_{\alpha\beta}^{(0)}=
g^{(0)}_{\alpha\beta}-\nabla_\alpha B_\beta -\nabla_\beta B_\alpha .
\label{eme}
\end{eqnarray}
In above equation, $B_\alpha $ is a vector field originating from
NS-NS two-form field $B_{(2)}$ in the K-S solution,
\begin{eqnarray}
B_\alpha = \partial_\alpha\left[ G \int _{S^2} B_{(2)}\right],
\label{sagoldb}
\end{eqnarray}
where $G$ is a certain function in ten dimensions. Since the
deformed $AdS_5$ vacuum configuration transforms as
\begin{eqnarray}
\delta \overline{g}^{(0)}_{\alpha\beta}=\nabla^{(0)}_\alpha
{\xi}_\beta + \nabla^{(0)}_\beta{\xi}_\alpha,
\end{eqnarray}
 the shifted background metric
$g^{(0)}_{\alpha\beta}$ is  invariant under the diffeomorphism
transformation,
\begin{eqnarray}
\delta x^\alpha=-\xi^\alpha, ~~~\delta B_\alpha=-{\xi}_\alpha,
\end{eqnarray}
Therefore, the $AdS_5$ space-time background is re-gained and
$SO(2,4)$ symmetry is recovered. A straightforward argument for this
vacuum configuration restoration is that $B_\alpha$ provides a
cancelation  to the non-symmetric part since $B_\alpha$ also has
$\ln (r/r_0)$ dependence.

 From Eqs.\,(\ref{reme}) and (\ref{eme}), the graviton in the
 symmetric
 $AdS_5$  background should be
\begin{eqnarray}
{h}_{\alpha\beta}= \overline{h}_{\alpha\beta}+\nabla^{(0)}_\alpha
B_\beta +\nabla^{(0)}_\beta B_\alpha . \label{sgold}
\end{eqnarray}
This shows that the graviton undergoes a gauge transformation from
which it picks up a longitudinal component and $B_\alpha$  plays the
role of a Goldstone vector field \cite{porra}.

Based on above analysis, we expand the classical  action
(\ref{t2bs}) to the second order of graviton $\overline{h}_{MN}$
around the symmetric $AdS_5$ background. Further, all of other
fields such as $B_{(2)}$ must be replaced by their $SO(2,4)$
transformed versions if they lie in certain non-trivial
representations of $SO(2,4)$. The quadratic action of type IIB
supergravity in deformed $AdS_5\times T^{1,1}$ background can be
recast into the one around the symmetric $AdS_5\times T^{1,1}$
vacuum configuration,
\begin{eqnarray}
\overline{S}_{\rm Q.G.} &=&\frac{1}{2k_{10}^2}\int d^{10}x
\sqrt{-G^{(0)}}\left\{-\frac{1}{4} \nabla^P \overline{h}^{MN}
\nabla_P \overline{h}_{MN}+\frac{1}{2}\nabla^P \overline{h}^{MN}
\nabla_M
\overline{h}_{PN}\right.\nonumber\\
&+&\frac{1}{4}\nabla^M \overline{h}\left( \nabla_M
\overline{h}-\frac{1}{2} \nabla^N \overline{h}_{MN}\right)
+\frac{27}{32}\frac{\pi \alpha^{\prime 2}N}{\overline{h}^2(r) r^5}
\left(\overline{h}^{MN}\overline{h}_{MN}- \frac{1}{2}h^2 \right)\nonumber\\
&+& \left.\frac{1}{2}\left[R^{(0)}_{MPNQ}-\frac{1}{90}
\left(G^{(0)}_{MN}G^{(0)}_{PQ}-G^{(0)}_{MQ}G^{(0)}_{NP}
\right)\frac{1}{2\times 3!} H^{(0)}_{(3)RST} H^{(0)RST}_{(3)}\right]
\overline{h}^{MN}\overline{h}^{PQ}\right\}\nonumber\\
&=&\frac{1}{2k_{10}^2}\int d^{10}x \sqrt{-G^{(0)}}\left[-\frac{1}{4}
\nabla^P \overline{h}^{MN}  \nabla_P
\overline{h}_{MN}+\frac{1}{2} \nabla_M \overline{h}^{MN} \nabla^P
\overline{h}_{PN}\right.\nonumber\\
&&+\frac{1}{4}\nabla^M \overline{h}\left( \nabla_M
\overline{h}-\frac{1}{2} \nabla^N
\overline{h}_{MN}\right)-\frac{9}{32}\frac{\pi \alpha^{\prime
2}N}{h^2(r) r^5}
\left(\overline{h}^{MN}\overline{h}_{MN}+ \frac{1}{2}\overline{h}^2 \right)\nonumber\\
&&\left.-\frac{1}{4}\times \frac{1}{2\times 45}\left(\frac{3g_s
M\alpha^\prime}{4} \right)^2 \frac{72}{r^6h^{3/2}(r)} \left(
\overline{h}^{MN} \overline{h}_{MN}-\overline{h}^2\right) \right].
\end{eqnarray}
The last term is the celebrated Pauli-Fierz mass term for type IIB
graviton \cite{pauli}.

Further, performing compactification on $T^{1,1}$ and taking into
account only zero modes in the K-K spectrum, we obtain the massive
$AdS_5$ graviton due to the spontaneous breaking of local $SO(2,4)$
symmetry to $SO(1,4)$,
\begin{eqnarray}
\overline{S}_{\rm q.g.} &=&\frac{1}{2k_{5}^2}\int d^{5}x
\sqrt{-g^{(0)}}\left[-\frac{1}{4} \nabla^\gamma
\overline{h}^{\alpha\beta} \nabla_\gamma
\overline{h}_{\alpha\beta}+\frac{1}{2}\nabla^\gamma
\overline{h}^{\alpha\beta} \nabla_\alpha
\overline{h}_{\gamma\beta}\right.\nonumber\\
&&+\frac{1}{4}\nabla^\alpha \overline{h}\left( \nabla_\alpha
\overline{h}-\frac{1}{2} \nabla^\beta
\overline{h}_{\alpha\beta}\right)-\frac{1}{L^2}
\left(\overline{h}^{\alpha\beta}\overline{h}_{\alpha\beta}+
\frac{1}{2}\overline{h}^2
\right)\nonumber\\
&&\left.-\frac{1}{4}m^2\left(\overline{h}^{\alpha\beta}
\overline{h}_{\alpha\beta}-\overline{h}^2\right) \right].
\label{afac11}
\end{eqnarray}
The mass $m$ can be evaluated by integrating over the internal
manifold $T^{1,1}$.

\subsection{Dual of $\gamma$-trace Anomaly $\gamma_\mu s^\mu$
for Supersymmetry Current} \label{subsect55}

Finally  we tackle the last member in the superconformal anomaly
multiplet, the dual of $\gamma$-trace anomaly  $\gamma^\mu s_\mu$ of
supersymmetry current $s_\mu$. As discussed in Sect.\,\ref{sect4},
the  breaking of super-Weyl symmetry  by fractional branes is
revealed by the Killing spinor equation obtained from the
supersymmetry transformation on the graviton $\Psi_M$ and dilatino
${\Lambda}$ in the K-S solution background.  The Killing spinor
equations in the symmetric and deformed $AdS_5\times T^{1,1}$
background are listed in Eq.\,(\ref{susytr2}) and
Eq.\,(\ref{susytr3}), respectively.
 We employ the same idea as
looking for the dual of scale anomaly to uncover the dual  of
$\gamma$-trace anomaly. That is, we should shift ${\Lambda}$ and
${\Psi}_M$ so that the Killing spinor equation (\ref{susytr3})  in
deformed $AdS_5\times T^{1,1}$ background recovers   the form of
Eq.\,(\ref{susytr2}), the Killing spinor equation in symmetric
$AdS_5\times T^{1,1}$ background.

A comparison between (\ref{susytr2}) and  (\ref{susytr3}) shows that
we should introduce a complex right-handed Weyl spinor $\Upsilon$
and make shift,
\begin{eqnarray}
{\Lambda}^\prime \equiv {\Lambda} -4i\Upsilon, ~~~ {\Psi}^\prime_M
\equiv \Psi_M-\Gamma_M \Upsilon. \label{fermionshift}
\end{eqnarray}
Using the property of $\Gamma$-matrix in ten dimensions,
\begin{eqnarray}
 \Gamma_{M_1M_2\cdots M_n}&=&\Gamma_{[M_1}\Gamma_{M_2}\cdots
\Gamma_{M_n]}=\frac{1}{n!}\sum_P
(-1)^{\delta_P}\Gamma_{a_{P(1)}}\Gamma_{a_{P(2)}}\cdots
\Gamma_{a_{P(n)}},\nonumber\\
\Gamma_{M_1M_2\cdots M_n N}&=& \Gamma_{M_1M_2\cdots M_n} \Gamma_N -
n\Gamma_{[M_1M_2\cdots M_{n-1}} G_{M_n] N},\nonumber\\
\Gamma_{NM_1M_2\cdots M_n }&=& \Gamma_N \Gamma_{M_1M_2\cdots M_n} -
nG_{N[M_1 } \Gamma_{M_1M_2\cdots M_{n}]} ,
\end{eqnarray}
we find that  $\Upsilon$ should be assigned to the supersymmetry
transformation
\begin{eqnarray}
\delta \Upsilon = -\frac{1}{96}\left({F}_{(3)MNP}+i{H}_{(3)MNP}
\right)\Gamma^{MNP} \epsilon
 \label{susytrans6}
\end{eqnarray}
so that (\ref{susytr2}) can be reproduced from (\ref{susytr3})
   (up to the linear term in fermionic fields and
 to the first order in the gravitational coupling $\kappa_{\rm 10}$). After the
 compactification, $\Upsilon$ will yield a Goldstone fermion needed
 for a super-Higgs mechanism, which should arise since
 the NS-NS- and R-R three-form fluxes breaks one-half of local
 supersymmetries.  Eq.\,(\ref{susytrans6}) implies that
 the supersymmetric transformation for the compactified $\Upsilon$ is
 proportional to the transformation parameter with the
 proportionality  coefficient given by the three-form fluxes
 passing through $S^3$.
 This is a typical feature of the Goldstone fermion \cite{wezu}.

 We focus on  the quadratic
 gravitino action of type IIB supergravity,
\begin{eqnarray}
S_{\rm gravitino}&=&\frac{1}{2k_{\rm 10}^2} \int
d^{10}x\,\sqrt{-G}\left[ -\frac{i}{2}\overline{\Psi}_M\Gamma^{MNP}
 D_N\Psi_P\right.\nonumber\\
 &&\left.-\frac{1}{8\times 5!}\overline{\Psi}_M \Gamma^{MNP}\left(
\Gamma^{UVWXY}{F}_{(5)UVWXY}\right)\Gamma_N\Psi_P+\cdots \right].
\label{qttb}
\end{eqnarray}
From Eq.\,(\ref{at11}), when the fractional D3-branes are absent,
the space-time background is $AdS_5 \times T^{1,1}$ with the
self-dual five-form field strength,
\begin{eqnarray}
F_{x_0 x_1 x_2 x_3 r}=\frac{r^3}{g_sL^4}, ~~~F_{g_1g_2 g_3g_4
g_5}=\frac{L^4}{27 g_s}. \label{selff}
\end{eqnarray}

In the following we perform the compactification of fermionic action
(\ref{qttb}) on $T^{1,1}$. With the  $\Gamma$-matrix representations
listed in Appendix,
 $\Psi_M$, ${\Lambda}$ and the supersymmetry
transformation parameter $\epsilon$
 decompose as the following,
\begin{eqnarray}
\Psi_\alpha (x,y) &=& \widehat{\psi} _\alpha (x,y)
\otimes \left(\begin{array}{c} 1\\
0\end{array}\right)= \left(\begin{array}{c} \widehat{\psi} _\alpha
(x,y)
\\
0\end{array}\right),\nonumber\\
\Psi_\xi (x,y) &=&\widehat{\psi} _\xi (x,y)
\otimes \left(\begin{array}{c} 1\\
0\end{array}\right)= \left(\begin{array}{c} \widehat{\psi} _\xi
(x,y)
\\
0\end{array}\right), \nonumber\\
\Lambda (x,y) &=& \widehat{\Lambda}(x,y)\otimes \left(\begin{array}{c} 0\\
1\end{array}\right)=\left(\begin{array}{c} 0\\
\widehat{\Lambda} (x,y)\end{array}\right), \nonumber\\
\epsilon (x,y)
&=&  \widehat{\epsilon} (x,y) \otimes \left(\begin{array}{c} 1\\
0\end{array}\right)=\left(\begin{array}{c} \widehat{\epsilon} (x,y)\\
0\end{array}\right),
\nonumber\\
\Upsilon (x,y) &=& \widehat{\chi}(x,y) \otimes \left(\begin{array}{c} 0\\
1\end{array}\right)=\left(\begin{array}{c} 0\\
\widehat{\chi} (x,y)\end{array}\right),
 \label{mwspinor}
\end{eqnarray}
where $\widehat{\psi}_\alpha$, $\widehat{\psi}_\xi$,
$\widehat{\Lambda}$, $\widehat{\epsilon}$ and $\widehat{\chi}$ are
the 16-component Weyl spinor in ten dimensions.  We then expand
these field functions and the supersymmetry transformation parameter
in the four-component $SO(5)$ spinor harmonics, which further
decompose into one-dimensional $U_{\rm H}(1)$-harmonics on $T^{1,1}$
\cite{ita}, that is,
\begin{eqnarray}
\widehat{\psi}_{\alpha} (x,y) &=&\sum_{j,l,r}
\psi_{\alpha}^{(j,l,r)}
(x)\Xi^{(j,l,r)}(y)=\sum_{j,l,r}\left(\begin{array}{c}
\psi^{(j,l,r-1)}_{\alpha} (x) Y^{(j,l,r-1)}_{0} (y)\\
\psi^{(j,l,r+1)}_{\alpha} (x) Y^{(j,l,r+1)}_{0} (y)\\
\psi^{(j,l,r)}_{\alpha} (x) Y^{(j,l,r)}_{-1} (y)\\
\psi^{(j,l,r)}_{\alpha} (x) Y^{(j,l,r)}_{+1} (y)
 \end{array}\right).
\label{spexpan}
\end{eqnarray}
$\widehat{\psi}_\xi (x,y)$, $\widehat{\Lambda} (x,y)$,
$\widehat{\epsilon} (x,y)$ and $\widehat{\chi} (x,y)$ admit the same
expansion in terms of the $U_{\rm H}(1)$-harmonics on $T^{1,1}$.
 Note that $\psi_\alpha^{(j,l,r)} (x)$, $\Lambda^{(j,l,r)} (x)$,
and $\epsilon^{(j,l,r)} (x)$ are four-component spinors on $AdS_5$;
$\Xi^{(j,l,r)} (y)$ are the four-component $SO(5)$ spinor harmonics
on $T^{1,1}$, and $Y^{(j,l,r)}_{q}(y)$ are $U_{\rm H}(1)$-harmonics
carrying $U_{\rm H}(1)$-charge $q$.

We substitute the chiral decomposition (\ref{mwspinor})
 and  the $U_{\rm H}(1)$-harmonic expansions for $\Psi_\alpha$
 and ${\Lambda}$ into the quadratic action (\ref{qttb})
 and consider only zero modes of their  kinetic operators defined
 on $T^{1,1}$.  Further, integrating over $T^{1,1}$
  we obtain
 the quadratic action for the graviton of the
 gauged $AdS_5$ supergravity,
\begin{eqnarray}
S_{\rm gravitino}&=&\frac{1}{2\kappa_{\rm 5}^2} \int
d^{10}x\,e\left(
-\frac{i}{2}\overline{\psi}^i_\alpha\gamma^{\alpha\beta\gamma}
 D_\beta\psi_{i\gamma} + \frac{3}{4 L}
  \overline{\psi}_\alpha^i \gamma^{\alpha\beta}
 \psi_{i\beta}
 +\cdots\right).
 \label{maslac}
\end{eqnarray}
The second term in Eq.\,(\ref{maslac}) is required by supersymmetry
to accompany the cosmological term in $AdS_5$ space \cite{send};
$\psi_\alpha^i$ is the $SU(2)$ symplectic Majorana spinor and the
index $i=1,2$ labels two gravitini  with $U(1)$-charge $r=\pm 1$,
which arise automatically when performing compactification on
$T^{1,1}$ \cite{ita}.

On the other hand, the fractional $D3$ branes deform the
$AdS_5\times T^{1,1}$ background. The gravitino action (\ref{qttb})
can be expressed in terms of either $\Psi_M$, ${\Lambda}$ in
deformed $AdS_5 \times T^{1,1}$ background or the shifted  fields
$\Psi_M^\prime$ and ${\Lambda}^\prime$ in the symmetric $AdS_5
\times T^{1,1}$ background. We start from the second  term of
(\ref{qttb}) in the deformed  $AdS_5\times T^{1,1}$ background and
re-write it in terms of shifted fermionic field variables
(\ref{fermionshift}) in the $AdS_5\times T^{1,1}$ background. Up to
the leading order of $\kappa_{\rm 10}$ we have
\begin{eqnarray}
&& \overline{S}_{\rm gravitino}=\frac{1}{2\kappa_{\rm 10}^2} \int
d^{10}x\,\sqrt{-\overline{G}}\left[ -\frac{1}{8\times
5!}\overline{\Psi}_M \Gamma^{MNP}\left(
\Gamma^{UVWXY}\overline{\widetilde{F}}_{(5)UVWXY}\right)
\Gamma_N\Psi_P\right]\nonumber\\
&=& \frac{1}{2\kappa_{\rm 10}^2} \int
d^{10}x\,\sqrt{-G}\left(-\frac{1}{8\times 5!} \right)\left(
\overline{\Psi}^\prime_M + \overline{\Upsilon}\Gamma_M
\right)\Gamma^{MNP}
\left(\Gamma^{(5)}\cdot\widetilde{F}_{(5)}\right)\Gamma_N \left(
{\Psi}_P^\prime- \Gamma_P {\Upsilon} \right) \nonumber\\
&=& \frac{1}{2\kappa_{\rm 10}^2} \int
d^{10}x\,\sqrt{-G}\left(-\frac{1}{8\times 5!} \right)\left(
\overline{\Psi}^\prime_M \Gamma^{MNP}+8
\overline{\Upsilon}\Gamma^{NP}\right)
\left(\Gamma^{(5)}\cdot\widetilde{F}_{(5)}\right)
 \left(\Gamma_N\Psi_P^\prime-\Gamma_{NP}\Upsilon\right)\nonumber\\
 &=& -\frac{1}{2\kappa_{\rm 10}^2}\frac{1}{8\times 5!} \int
d^{10}x\,\sqrt{-G}\left[ \left(\overline{\Psi}_\alpha^\prime
\Gamma^{\alpha\beta\gamma}+\overline{\Psi}_\xi^\prime \Gamma^\xi
\Gamma^{\beta\gamma}+8
\overline{\Upsilon}\Gamma^{\beta\gamma}\right)\right.\nonumber\\
&&\times \left(
\Gamma^{\alpha_1\cdots\alpha_5}\widetilde{F}_{\alpha_1\cdots\alpha_5}
+\Gamma^{\xi_1\cdots\xi_5}\widetilde{F}_{\xi_1\cdots\xi_5}
\right)\left(\Gamma_\beta\Psi_\gamma^\prime-\Gamma_{\beta\gamma}\Upsilon
\right)
\nonumber\\
&+&\left(\overline{\Psi}_\alpha^\prime
\Gamma^{\alpha}\Gamma^{\xi\varepsilon}+\overline{\Psi}_\tau^\prime
\Gamma^{\tau\xi\varepsilon}+8
\overline{\Upsilon}\Gamma^{\xi\varepsilon}\right) \left(
\Gamma^{\alpha_1\cdots\alpha_5}\widetilde{F}_{\alpha_1\cdots\alpha_5}
+\Gamma^{\xi_1\cdots\xi_5}\widetilde{F}_{\xi_1\cdots\xi_5}
\right)\left(\Gamma_\xi\Psi_\varepsilon^\prime
-\Gamma_{\xi\varepsilon}\Upsilon
\right)\nonumber\\
&+&\left(\overline{\Psi}_\alpha^\prime
\Gamma^{\alpha\beta}\Gamma^{\xi}+\overline{\Psi}_\varepsilon^\prime
\Gamma^{\varepsilon\xi}\Gamma^\beta +8
\overline{\Upsilon}\Gamma^{\beta}\Gamma^{\xi}\right)\left(
\Gamma^{\alpha_1\cdots\alpha_5}\widetilde{F}_{\alpha_1\cdots\alpha_5}
+\Gamma^{\xi_1\cdots\xi_5}\widetilde{F}_{\xi_1\cdots\xi_5}
\right)\left(\Gamma_\beta\Psi_\xi^\prime-\Gamma_{\beta}\Gamma_{\xi}\Upsilon
\right)\nonumber\\
&+&\left.\left(-\overline{\Psi}_\alpha^\prime
\Gamma^{\alpha\beta}\Gamma^{\xi}+\overline{\Psi}_\varepsilon^\prime
\Gamma^{\varepsilon\xi}\Gamma^\beta + 8
\overline{\Upsilon}\Gamma^{\beta}\Gamma^{\xi}\right)\left(
\Gamma^{\alpha_1\cdots\alpha_5}\widetilde{F}_{\alpha_1\cdots\alpha_5}
+\Gamma^{\xi_1\cdots\xi_5}\widetilde{F}_{\xi_1\cdots\xi_5}
\right)\left(\Gamma_\xi\Psi_\beta^\prime-\Gamma_{\xi}\Gamma_{\beta}\Upsilon
\right)\right]\nonumber\\
&=& -\frac{1}{2\kappa_{\rm 10}^2}\frac{1}{ 5!} \int
d^{10}x\,\sqrt{-G}\left[\frac{1}{4}\overline{\Psi}_\alpha^\prime
\Gamma^{\alpha\beta}
\left(\Gamma^{\alpha_1\cdots\alpha_5}\widetilde{F}_{\alpha_1\cdots\alpha_5}
-\Gamma^{\alpha_1\cdots\alpha_5}\widetilde{F}_{\alpha_1\cdots\alpha_5}
\right)\Psi_\beta^\prime \right.\nonumber\\
&&+\overline{\Psi}_\alpha^\prime
\Gamma^{\alpha}\Gamma^{\xi}\left(\Gamma^{\alpha_1\cdots\alpha_5}
\widetilde{F}_{\alpha_1\cdots\alpha_5}
-\Gamma^{\xi_1\cdots\xi_5}\widetilde{F}_{\xi_1\cdots\xi_5}
\right)\Psi_\xi^\prime\nonumber\\
&&+ \overline{\Upsilon}\Gamma^\alpha
\left(\Gamma^{\alpha_1\cdots\alpha_5}\widetilde{F}_{\alpha_1\cdots\alpha_5}
-\Gamma^{\alpha_1\cdots\alpha_5}\widetilde{F}_{\alpha_1\cdots\alpha_5}
\right)\Psi_\alpha^\prime-\overline{\Psi}^\prime_\alpha\Gamma^\alpha
\left(\Gamma^{\alpha_1\cdots\alpha_5}\widetilde{F}_{\alpha_1\cdots\alpha_5}
+\Gamma^{\xi_1\cdots\xi_5}\widetilde{F}_{\xi_1\cdots\xi_5}
\right)\Upsilon\nonumber\\
&&- \overline{\Psi}_\xi^\prime\Gamma^\xi
\left(\Gamma^{\alpha_1\cdots\alpha_5}\widetilde{F}_{\alpha_1\cdots\alpha_5}
+\Gamma^{\xi_1\cdots\xi_5}\widetilde{F}_{\xi_1\cdots\xi_5}
\right)\Upsilon+\overline{\Upsilon}\Gamma^\xi
\left(\Gamma^{\alpha_1\cdots\alpha_5}\widetilde{F}_{\alpha_1\cdots\alpha_5}
-\Gamma^{\xi_1\cdots\xi_5}\widetilde{F}_{\xi_1\cdots\xi_5}
\right)\Psi_\xi^\prime\nonumber\\
&&-2\times5 \times 5\overline{\Upsilon}\,\Gamma^\alpha
\left(\Gamma^{\alpha_1\cdots\alpha_5}\widetilde{F}_{\alpha_1\cdots\alpha_5}
+\Gamma^{\xi_1\cdots\xi_5}\widetilde{F}_{\xi_1\cdots\xi_5}
\right)\Upsilon\nonumber\\
&&\left.-\frac{1}{4}\,\overline{\Psi}_\xi^\prime
\Gamma^{\xi\varepsilon}\left(\Gamma^{\alpha_1\cdots\alpha_5}
\widetilde{F}_{\alpha_1\cdots\alpha_5}
-\Gamma^{\xi_1\cdots\xi_5}\widetilde{F}_{\xi_1\cdots\xi_5}
\right)\Psi_\varepsilon^\prime \right], \label{expgraviton}
\end{eqnarray}
where the following operation is performed,
\begin{eqnarray}
\left[\Gamma^\alpha,\Gamma^{\alpha_1\cdots\alpha_5}
\right]\widetilde{F}_{\alpha_1\cdots\alpha_5}&=&\left(5 g^{\alpha
[\alpha_1}\Gamma^{\alpha_2\cdots\alpha_5]}-5
\Gamma^{[\alpha_1\cdots\alpha_4}g^{\alpha_5]\alpha}
\right)\widetilde{F}_{\alpha_1\cdots\alpha_5} \nonumber\\
&=&
5g^{\alpha\alpha_1}\Gamma_{\alpha_1}-5g^{\alpha\alpha_1}\Gamma_{\alpha_1}=0,
\nonumber\\
\left[\Gamma^\xi,\Gamma^{\xi_1\cdots\xi_5}
\right]\widetilde{F}_{\xi_1\cdots\xi_5}
&=&\left[\Gamma^{\xi\varepsilon},\Gamma^{\xi_1\cdots\xi_5}
\right]\widetilde{F}_{\xi_1\cdots\xi_5}
=\left[\Gamma^{\alpha\beta},\Gamma^{\alpha_1\cdots\alpha_5}
\right]\widetilde{F}_{\alpha_1\cdots\alpha_5}=0.
\end{eqnarray}
Further,  the explicit $\Gamma$-matrix representation listed in
(\ref{gmat1}) and (\ref{gmat2}) gives
\begin{eqnarray}
&&\Gamma^{\alpha_1\cdots\alpha_5}=-i\epsilon^{\alpha_1\cdots\alpha_5}1_4\otimes
1_4\otimes \sigma_1,~~~
\Gamma^{\xi_1\cdots\xi_5}=-\epsilon^{\xi_1\cdots\xi_5}1_4\otimes
1_4\otimes \sigma_2,\nonumber\\
&&
\Gamma^{\alpha_1\cdots\alpha_5}\widetilde{F}_{\alpha_1\cdots\alpha_5}
\pm
\Gamma^{\xi_1\cdots\xi_5}\widetilde{F}_{\xi_1\cdots\xi_5}\nonumber\\
&=&-i 1_4\otimes 1_4\otimes \left( \begin{array}{cc} 0 &
\epsilon^{\alpha_1\cdots\alpha_5}\widetilde{F}_{\alpha_1\cdots\alpha_5}\mp
\epsilon^{\xi_1\cdots\xi_5}\widetilde{F}_{\xi_1\cdots\xi_5}
\\
\epsilon^{\alpha_1\cdots\alpha_5}\widetilde{F}_{\alpha_1\cdots\alpha_5}\pm
\epsilon^{\xi_1\cdots\xi_5}\widetilde{F}_{\xi_1\cdots\xi_5} & 0
\end{array} \right).
\label{gacal}
\end{eqnarray}
Substituting (\ref{gacal}) and the Weyl spinor representations
(\ref{mwspinor}) into (\ref{expgraviton}), we  reduce
$\overline{S}_{\rm gravitino}$ to the form expressed in the
ten-dimensional Weyl spinors,
\begin{eqnarray}
\overline{S}_{\rm gravitino}&=&\frac{1}{2\kappa_{\rm 10}^2}\frac{i}{
5!} \int
d^{10}x\,\sqrt{-G}\left[\left(\frac{1}{4}\overline{\widehat{\psi}}_\alpha
\gamma^{\alpha\beta}\widehat{\psi}_\beta-2\overline{\widehat{\psi}}_\alpha
\gamma^\alpha \widehat{\chi}+5 \overline{\widehat{\chi}}
\widehat{\chi} \right)\right.\nonumber\\
&&\times \left.
\left(\epsilon^{\alpha_1\cdots\alpha_5}\widetilde{F}_{\alpha_1\cdots\alpha_5}
-\epsilon^{\xi_1\cdots\xi_5}\widetilde{F}_{\xi_1\cdots\xi_5}
\right)\right]\nonumber\\
&=&-\frac{i}{2\kappa_{\rm 10}^2} \int
d^{10}x\,\sqrt{-G}\,\left[\frac{3}{L}\,\frac{1}{g_s}\left(1+\frac{3}{2\pi}\frac{g_s
M^2}{N}\ln\frac{r}{r_0}\right)\right.\nonumber\\
&&\left.\times \left(\frac{1}{4}\overline{\widehat{\psi}}_\alpha
\gamma^{\alpha\beta}\widehat{\psi}_\beta-2\overline{\widehat{\psi}}_\alpha
\gamma^\alpha \widehat{\chi}+5 \overline{\widehat{\chi}}
\widehat{\chi} \right)\right]. \label{superhi}
\end{eqnarray}
In above derivation we have used (\ref{selff}),
\begin{eqnarray}
&& \frac{1}{5!}\left(\epsilon^{\alpha_1\cdots\alpha_5}
\widetilde{F}_{\alpha_1\cdots\alpha_5} -\epsilon^{\xi_1\cdots\xi_5}
\widetilde{F}_{\xi_1\cdots\xi_5}\right)\nonumber\\
&=&-(-g_{AdS_5})^{-1/2}F_{0123r}-(g_{
T^{1,1}})^{-1/2}F_{g^1g^2g^3g^4g^5} \nonumber\\
&=&-3 h^{-5/4}(r)\frac{27}{4}\frac{\pi \alpha^{\prime 2}N_{\rm
eff}}{r^5}
=-\frac{3}{L}\,\frac{1}{g_s}\left(1+\frac{3}{2\pi}\frac{g_s
M^2}{N}\ln\frac{r}{r_0} \right).
\end{eqnarray}

Eq.\,(\ref{superhi}) presents the typical feature of super-Higgs
effects in supergravity \cite{wezu} with $\widehat{\chi}$ playing
the role of a Goldstone fermion. Substituting the $U_{\rm
H}(1)$-harmonic expansions (\ref{spexpan})
 for $\widehat{\psi}_\alpha$ and
$\widehat{\chi}$ into (\ref{superhi}), taking into account only zero
modes, and further integrating over the internal manifold $T^{1,1}$,
we obtain
\begin{eqnarray}
{S}_{\rm gravitino}=-\frac{i}{2\kappa_{\rm 5}^2} \int
d^{5}x\,\sqrt{-g^{(0)}}\,\left(\frac{3}{4L}\,\frac{1}{g_s}+m\right)
\left(\overline{\psi}^i_\alpha
\gamma^{\alpha\beta}{\psi}_{i\beta}-8\overline{\psi}^i_\alpha
\gamma^\alpha {\chi}_i+20 \overline{\chi}^i{\chi}_i \right)
\label{superhi2}
\end{eqnarray}
Finally,  with a shift $\psi^{i\prime}_\mu =\psi^i_\mu-\gamma_\mu
\chi^i$, an  elegant result appears,
\begin{eqnarray}
{S}_{\rm gravitino} &=&-\frac{i}{2\kappa_{\rm 5}^2} \int
d^{5}x\,\sqrt{-g^{(0)}}\,\left(\frac{3}{4L}\,\frac{1}{g_s}+m\right)
\overline{\psi}_\alpha^{i\prime}
\gamma^{\alpha\beta}{\psi}_\beta^{i\prime}. \label{superhi3}
\end{eqnarray}
 The first term with  coefficient
proportional to $1/(Lg_s)$ in (\ref{superhi3})  accompanies the
cosmological constant term, which is required by the Poincar\'{e}
supersymmetry \cite{send}; the second one is the mass term for the
gravitino in five-dimensional gauged supergravity. This mass is
generated  by super-Higgs mechanism and is proportional to fluxes
$M$ carried by fractional branes, which can be seen clearly from
Eq.\,(\ref{superhi}).

\subsection{Goldstone Hypermultiplet in $AdS_5$ Space
%${\cal N}=1$ Goldstone Chiral Supermultiplet on $AdS_5$ Boundary
} \label{subsect56}

We have shown  that  the superconformal anomaly multiplet of an
${\cal N}=1$ $SU(N+M)\times SU(N)$  supersymmetric gauge theory in
four dimensions is dual to the spontaneous breaking of local
supersymmetry and the consequent super-Higgs mechanism in  ${\cal
N}=2$ $U(1)$ gauged $AdS_5$ supergravity, through which the ${\cal
N}=2$ graviton supermultiplet
($h_{\alpha\beta}$,$\psi^i_\alpha$,$A_\alpha$) becomes massive. A
crucial ingredient in implementing this super-Higgs mechanism is the
Goldstone fields $\theta$, $B_\alpha$ and $\Upsilon$ in ten
dimensions, which are defined in Eqs.\,(\ref{fgold}),
(\ref{sagoldb}), (\ref{fermionshift}) and (\ref{susytrans6}),
respectively. Since the superconformal anomaly in an ${\cal N}=1$
four-dimensional supersymmetric gauge theory is an ${\cal N}=1$
chiral supermultiplet \cite{grisa}, so according to the AdS/CFT
correspondence conjecture \cite{mald,gkp,witt1}, $\theta$,
$B_\alpha$ and $\Upsilon$ should constitute a supermultiplet.
Specifically, the holographic version on AdS/CFT correspondence at
supergravity level \cite{gkp,witt1} requires that after the
compactification on $T^{1,1}$ $\theta$, $B_\alpha$ and $\Upsilon$
should form an ${\cal N}=2$ Goldstone supermultiplet
(hypermultiplet) in $AdS_5$ space and its $AdS_5$ boundary value
should be an ${\cal N}=1$ chiral supermultiplet in four dimensions.
In the following we verify this fact.

As a first step, we start from the supersymmetric transformations
for $B_{(2)MN}$ and $F_{(3)MNP}$ \cite{schw},
\begin{eqnarray}
\delta G_{(3)MNP}&=& 3\partial_{[M}\delta A_{(2)NP]}=\delta
\left[F_{(3)MNP}+iH_{(3)MNP}\right]\nonumber\\
&=& 3\partial_{[M}\delta
C_{(2)NP]}+3i \partial_{[M}\delta B_{(2)NP]}\nonumber\\
&=&
3\partial_{[M}\left(\overline{\epsilon}\Gamma_{NP]}{\Lambda}\right)
-12i\partial_{[M}\left(\overline{\epsilon}^\star \Gamma_N
\Psi_{P]}\right).
\end{eqnarray}
 This shows  that locally there should exist
\begin{eqnarray}
 \delta A_{(2)NP} = \delta
C_{(2)NP}+i \delta B_{(2)NP} =
\overline{\epsilon}\Gamma_{NP}{\Lambda} -2i\overline{\epsilon}^\star
\left(\Gamma_N \Psi_{P}-\Gamma_P\Psi_N\right).
\end{eqnarray}
 When we restore the deformed $AdS_5\times T^{1,1}$ back to the symmetric
 $AdS_5\times T^{1,1}$
 vacuum background,
   the fermionic fields $\Psi_M$ and ${\Lambda}$
   are reparametrized as in (\ref{fermionshift}),
   the supersymmetric transformation for shifted $A_{(2)NP}$ becomes
\begin{eqnarray}
 \delta \overline{A}_{(2)NP} &=& \delta
\overline{C}_{(2)NP}+i \delta \overline{B}_{(2)NP}\nonumber\\
&=&
\overline{\epsilon}\Gamma_{NP}\left(\widehat{\lambda}-4i\Upsilon\right)
-2i\overline{\epsilon}^\star \left[\Gamma_N
\left(\Psi_{P}-\Gamma_P\Upsilon\right)
-\Gamma_P\left(\Psi_N-\Gamma_P\Upsilon\right)\right]\nonumber\\
&=&\overline{\epsilon}\Gamma_{NP}\widehat{\lambda}
-2i\overline{\epsilon}^\star \left(\Gamma_N
\Psi_{P}-\Gamma_P\Psi_N\right)-4i \left(
\overline{\epsilon}-\overline{\epsilon}^\star\right)
\Gamma_{NP}\Upsilon.
\end{eqnarray}
This leads to the  supersymmetric transformation involving the
Goldstone spinor $\Upsilon$,
\begin{eqnarray}
\delta {A}_{(2)NP}^{(\Upsilon)}=-4i \left(
\overline{\epsilon}-\overline{\epsilon}^\star\right)
\Gamma_{NP}\Upsilon =4 \left(\mbox{Im}\,\overline{\epsilon}\right)
\,\Gamma_{NP}\Upsilon, \label{goldsusy1}
\end{eqnarray}
where we use ${A}_{(2)NP}^{(\Upsilon)}$ to represent the parts of
shifted  $B_{(2)MN}$ and $C_{(2)MN}$ whose supersymmetric
transformations depend only on $\Upsilon$.

We turn to  the supersymmetric transformation (\ref{susytrans6}) for
$\Upsilon$,
\begin{eqnarray}
\delta\Upsilon &=&-\frac{1}{96}\times 3\,\partial_{[M}
A^{(\Upsilon)}_{(2)NP]}\Gamma^{MNP}\epsilon \nonumber\\
&=&-\frac{1}{96}\times 3\left(\partial_{[M}
C^{(\Upsilon)}_{(2)NP]}+i\partial_{[M}B^{(\Upsilon)}_{(2)NP]}
\right)\Gamma^{MNP}\epsilon\nonumber\\
&=&-\frac{1}{32}\Gamma^M\left(  \partial_M
A^{(\Upsilon)}_{(2)NP}\right)\Gamma^{NP}\epsilon, \label{goldsusy2}
\end{eqnarray}
where the following $\Gamma$-matrix relation is employed,
\begin{eqnarray}
\Gamma^{MNP}=\Gamma^M\Gamma^{NP}-
\left(G^{MN}\Gamma^P-G^{MP}\Gamma^N \right).
\end{eqnarray}
Eqs.\,(\ref{goldsusy1}) and (\ref{goldsusy2}) show that
$A^{(\Upsilon)}_{(2)NP}$ and $\Upsilon$ constitute a supermultiplet
in ten dimensions since their supersymmetric transformations form a
closed algebra.

Further, from the definitions (\ref{fgold}) and (\ref{sgold}) on the
Goldstone bosons,
\begin{eqnarray}
\theta = F\int _{S^2} C_{(2)}, ~~~B_\alpha=\partial_\alpha \left[
G\int_{S^2} B_{(2)}\right]\equiv \partial_\alpha \omega,
\end{eqnarray}
we choose $F=G$ and define a new complex scalar field in ten
dimensions,
\begin{eqnarray}
\widehat{\phi} \equiv \theta +i \omega  =F\left[\int _{S^2} \left(
C_{(2)}+i B_{(2)}\right)\right] \label{comsca}
\end{eqnarray}
It should be emphasized that the choice $F=G$ is always possible
since in (\ref{fgold}) and (\ref{sgold}) $F$ and $G$ are introduced
as arbitrary field functions in ten dimensions. Actually, the
requirement that the Goldstone fields should constitute a
supermultiplet imposes this choice. Furthermore, since in the K-S
solution both $C_{(2)}$ and $B_{(2)}$ are proportional to
$\omega_2=\left( g^1\wedge g^2+g^3\wedge g^4 \right)/2$, the complex
scalar field $\widehat{\phi}$ defined in (\ref{comsca}) actually
contains two complex (four real) scalar fields,
\begin{eqnarray}
\widehat{\phi}^1 \sim  \int_{S^2}\left( C_{(2)g_1g_2}+i
B_{(2)g_1g_2}\right), ~~~ \widehat{\phi}^2 \sim  \int_{S^2}\left(
C_{(2)g_3g_4}+i B_{(2)g_3g_4}\right).
\end{eqnarray}
Consequently, the supersymmetric transformations (\ref{goldsusy1})
and (\ref{goldsusy2}) for the Goldstone multiplet
$(A^{(\Upsilon)}_{(2)NP},\Upsilon)$ take the following form,
\begin{eqnarray}
\delta \widehat{\phi}^1 &=&
4\left(\mbox{Im}\,\overline{\epsilon}\right) \,
\Gamma_{g_1g_2}\widetilde{\Upsilon}, ~~~ \delta \widehat{\phi}^2 =
4\left(\mbox{Im}\,\overline{\epsilon}\right) \,
\Gamma_{g_3g_4}\widetilde{\Upsilon}, \nonumber\\
\delta \widetilde{\Upsilon} &=& -\frac{1}{32}\Gamma^M \left[
\left(\partial_M \phi^1\right)\Gamma_{g_1g_2}+ \left(\partial_M
\phi^2\right)\Gamma_{g_3g_4}\right]{\epsilon}, ~~~
\widetilde{\Upsilon} \equiv \int_{S^2}\Upsilon . \label{goldsusy3}
\end{eqnarray}

Recall that  $\Upsilon$ and its supersymmetric transformation
(\ref{susytrans6}) are introduced to counter the supersymmetric
transformation in the K-S solution background so that the Killing
spinor equation (\ref{susytr3})  can recover to the Killing spinor
equation (\ref{susytr2}) in  $AdS_5\times T^{1,1}$ background.
Therefore, the supersymmetry transformation parameter $\epsilon$ in
(\ref{goldsusy1}), (\ref{goldsusy2}) and  (\ref{goldsusy3}) should
satisfy the constraint equations (\ref{kicon2}). Because we use the
UV (large-$\tau$) limit of the K-S solution, i.e., the K-T solution,
Eq.\,(\ref{kicon2}) gives
\begin{eqnarray}
\Gamma_{g_1g_2}\epsilon=-\Gamma_{g_3g_4}\epsilon=\left.
i\left(\frac{\sinh\,\tau}{\cosh\,\tau}-\frac{1}{\cosh\,\tau}
\Gamma_{g_1g_3}\right)\epsilon\right|_{\tau\to\infty}=i\epsilon.
\label{goldsusy4}
\end{eqnarray}
So the supersymmetric transformation (\ref{goldsusy3}) of the
Goldstone multiplet reduces to an elegant form,
\begin{eqnarray}
\delta \widehat{\phi}^1 &=&
-4i\,\left(\mbox{Im}\,\overline{\epsilon}\right)
\,\widetilde{\Upsilon}, ~~~ \delta \widehat{\phi}^2 =
4i\,\left(\mbox{Im}\,\overline{\epsilon}\right) \,
\widetilde{\Upsilon}, \nonumber\\
\delta \widetilde{\Upsilon} &=& -\frac{i}{32}\Gamma^M
\left(\partial_M \widehat{\phi}^1- \partial_M
\widehat{\phi}^2\right){\epsilon}. \label{goldsusy5}
\end{eqnarray}

Substituting the chiral decomposition of $\Upsilon$ and $\epsilon$
listed in (\ref{mwspinor}) into (\ref{goldsusy5}) and using
\begin{eqnarray}
\overline{\epsilon}={\epsilon}^\dagger \Gamma_0=
\left(\widehat{\epsilon}^\dagger  (x,y), 0\right) \, \gamma^0\otimes
1_4\otimes\left( \begin{array}{cc} 0 & 1 \\ 1 & 0 \end{array}
\right) =\left(0, \overline{\widehat{\epsilon}} (x,y) \right)
\end{eqnarray}
 we obtain
\begin{eqnarray}
\delta \widehat{\phi}^1 (x,y) &=& -4i
\left(\mbox{Im}\,\overline{\widehat{\epsilon}}\right) \,
\widehat{\widetilde{\chi}}(x,y), ~~~ \delta \widehat{\phi}^2 (x,y) =
4i \left(\mbox{Im}\,\overline{\widehat{\epsilon}}\right) \,
\widehat{\widetilde{\chi}}(x,y), \nonumber\\
\delta \widehat{\widetilde{\chi}}(x,y) &=& -\frac{i}{32}
\left[\gamma^\alpha \partial^x_\alpha
\left(\widehat{\phi}^1-\widehat{\phi}^2\right)
-i\tau^\xi\partial^y_\xi
\left(\widehat{\phi}^1-\widehat{\phi}^2\right)\right]\widehat{\epsilon}(x,y).
\label{goldsusy6}
\end{eqnarray}

The following task is performing compactification on $T^{1,1}$ and
reducing above supersymmetric transformations to $AdS_5$ space. We
expand $\widehat{\phi}^p(x,y)$ ($p=1,2$),
$\widehat{\widetilde{\chi}}(x,y)$ and $\widehat{\epsilon}(x,y)$ in
terms of the scalar- and spinor $U_H(1)$ harmonics as shown in
(\ref{spexpan}), and substitute the expansions into
Eq.\,(\ref{goldsusy6}). Taking into account only zero modes in the
expansions  and comparing both sides of Eq.\,(\ref{goldsusy6}), we
obtain
\begin{eqnarray}
\delta \phi^p = 4i f^p_{ij}\overline{\eta}^i \chi^j, ~~~~ \delta
\chi^i =-\frac{1}{32}if^{ij}_p\gamma^\alpha
\partial_\alpha
\phi^p\eta_j.
 \label{bulksusy}
\end{eqnarray}
In (\ref{bulksusy}), $\phi^p$, $\chi^i$ and $\eta^i$ are the scalar
fields, fermionic fields and ${\cal N}=2$ supersymmetry
transformation parameters in five dimensions obtained from the
compactification of $\widehat{\phi}^p$, $\widehat{\widetilde{\chi}}$
and $\widehat{\epsilon}$, respectively; $i,j=1,2$ are the indices
labeling fermions with opposite $U(1)$-charges $r=\pm 1$, which
arise naturally in performing the compactification of fermionic
fields on $T^{1,1}$,
 $f^{ij}_p\equiv\delta^{ij} N^{j p}$ ($p$ is not
summed); $N^{\upsilon A}$ are defined as the following:
$N^{11}=N^{21}=1$ and $N^{12}=N^{22}=-1$; $f^p_{ij}$ is related to
$f_p^{ij}$ as the following, $f^p_{ij}\equiv
\epsilon^{pq}\delta_{ik}\delta_{jl}f^{kl}_q$. Eq.\,(\ref{bulksusy})
is exactly the supersymmetry transformation for ${\cal N}=2$
hypermultiplet in five-dimensions \cite{gstow}.

We should further take the hypermultiplet $(\phi^p,\chi^i)$ to the
boundary of $AdS_5$ space. In principle, we can take a similar
procedure as in Ref.\,\cite{moda}, where on-shell fields in gauged
$AdS_{p+1}$ supergravity can reduce to the off-shell fields of
$p$-dimensional  conformal supergravity on $AdS_{p+1}$ boundary.
However, in the case at hand the
 equations of  motion for $\phi^p$ and $\chi^i$ are not clear.
 We naively take them on the boundary of $AdS_5$ space
 and assume rudely that Eq.\,(\ref{bulksusy})  should lead to
 the supersymmetry transformations  for  ${\cal N}=1$ chiral
supermultiplet in four dimensions. This feature corresponds  to the
superconformal anomaly in a four-dimensional ${\cal N}=1$
supersymmetric gauge theory.

\section{Summary}
\label{sect6}

We have investigated the supergravity dual description to the
superconformal anomaly  of a four-dimensional supersymmetric gauge
theory. We make use of two distinct properties of $D$-brane in type
II superstring theory: On one hand, in the weak coupling case, it
behaves as a dynamical and geometric object with open strings ending
on it. Thus a supersymmetric gauge theory on the world-volume of a
stack of coincident  $D$-branes can be constructed by an appropriate
construction on brane configuration, and hence all the quantum
phenomena of a gauge theory can be extracted out from $D$-brane
dynamics; On the other hand, a $D$-brane is charged with respect to
the R-R string states and hence a stack  of $D$-branes modify the
space-time background of type II superstring theory. This type of
effect of $D$-branes make them behave as the brane solution to type
II supergravity in a strongly coupled type II superstring theory. We
specialize to ${\cal N}=1$ $SU(N+M)\times SU(N)$ Yang-Mills theory
with two matter fields in the bifundamental representations
$(N+M,\overline{N})$ and $(\overline{N+M},N)$ of gauge groups and a
quartic superpotential.  The brane configuration producing this
supersymmetric gauge theory consists of $N$ bulk $D3$-branes and $M$
fractional $D3$-branes in the singular target space-time $M^4\times
{\cal C}_6$.  The fractional branes are fixed at the apex of the
conifold ${\cal C}^6$ whose base is the five-dimensional Einstein
manifold $T^{1,1}\sim S^2\times S^3$. On the other hand, the
space-time background  arising from the brane configuration is the
celebrated Klebanov-Strassler solution. It is a non-singular
solution to type IIB supergravity and its UV limit can be considered
a deformed $AdS_5\times T^{1,1}$. We start from the field theory
result on superconformal anomaly and track its origin to the
$D$-brane configuration. We realize that the fractional branes
frozen are the origin for superconformal anomaly: When fractional
branes are absent, the field theory is an ${\cal N}=1$ $SU(N)\times
SU(N)$ supersymmetric gauge theory with two flavors in the
bifundamental representation $(N,\overline{N})$ and
$(\overline{N},N)$, which is a
 superconformal quantum gauge theory at the IR fixed point of its
renormalization group flow; While when fractional branes switch on,
the field theory becomes an ${\cal N}=1$ $SU(N+M)\times SU(N)$
supersymmetric gauge theory with two flavors in the bifundmental
representation $(N+M,\overline{N})$ and $(\overline{N+M},N)$, which
ceases to be a superconformal theory. Then we go to the strongly
coupled side of type II superstring and make use of the
gravitational feature of $D$-branes. We find that the effect of the
fractional $D3$-branes is to deform $AdS_5\times T^{1,1}$
space-time: Without fractional branes the near-horizon limit of
three-brane solution yielded from above brane configuration is
$AdS_5\times T^{1,1}$; While with the fractional branes present, the
corresponding three-brane solution transits to the K-S solution. We
choose the (UV limit of) K-S solution as a vacuum configuration for
type IIB supergravity. Due to the geometric meaning of a gravity
theory, the spontaneous compactification on the deformed $T^{1,1}$
takes place. Since in the Kaluza-Klein compactification, the
isometry symmetry of the background space-time and the supersymmetry
it preserves convert into local symmetries for the compactified
theory, so the type IIB supergravity in the $AdS_5\times T^{1,1}$
background should lead to ${\cal N}=2$ $U(1)$ gauged $AdS_5$
supergravity coupled with $SU(2)\times SU(2)$ Yang-Mills vector
multiplets and some Betti multiplets. While  in the K-S solution
background the type IIB supergravity  should yield a gauged $AdS_5$
supergravity with local symmetry breaking since the deformed
$AdS_5\times T^{1,1}$ is less symmetric. The broken symmetries
include the $U(1)$ gauge symmetry, ${\cal N}=1$ conformal
supersymmetry and part of diffeomorphism symmetry of $AdS_5$ space.
Further, the spontaneous breaking of local symmetry triggers the
Higgs mechanism. We work out the super-Higgs effect corresponding to
the spontaneous breaking of the above local symmetries and  show
that the Goldstone fields come from the NS-NS-  and R-R two-form
fields relevant to fractional branes. Finally, we verify that the
Goldstone fields, which is a key ingredient in implementing the
super-Higgs mechanism, do constitute an ${\cal N}=2$ hypermultiplet
in $AdS_5$ space and further we argue that it may lead to an ${\cal
N}=1$ chiral supermultiplet on the boundary of $AdS_5$ space. Since
the superconformal anomaly of a four-dimensional supersymmetric
gauge theory is a chiral supermultiplet, this verification coincides
with the prediction from gauge/gravity duality.

The presence of fractional $D3$-branes is the common origin for the
superconformal anomaly of the supersymmetric gauge theory and the
spontaneous breaking of local symmetries in ${\cal N}=2$ $U(1)$
gauged $AdS_5$ supergravity. Therefore, we conclude that a dual
correspondence between the superconformal anomaly on the field
theory side and the super-Higgs mechanism on the gravity should be
established.

\acknowledgments I would like to thank Professor  M. Chaichian and
Professor R. Jackiw for their  continuous encouragements and support
and O. Aharony, A. Buchel, C.T. Chan, M. Grisaru, A. Kobakhidze, C.
N\'{u}\~{n}ez and Y. Yang for discussions, communications and
conversations. I am especially indebted to Professors I.R. Klebanov,
M. G\"{u}naydin and Dr. P. Ouyang for discussions in revising the
original version. I would also like to thank my colleagues at the
string theory group of Taiwan for comments when I gave several
seminars on this topic. This work is partially supported by the
National Research Council through NCTS of Taiwan under the grant NSC
94-2119-M-002-001.

\appendix

\section{ $\Gamma$-matrix in  ten dimensions}

We choose following explicit representation for $\Gamma$-matrices
$\Gamma^A$ in ten dimensions \cite{ita,kim},
\begin{eqnarray}
\Gamma^A &=& e^A_{~~M}\Gamma^M, ~~\Gamma^a = \gamma^a \otimes 1_4
\otimes \sigma^1, ~~\Gamma^m= 1_4 \otimes \tau^m \otimes
(-\sigma^2),  \label{gmat1}
\end{eqnarray}
where $M= (\alpha,\xi)$ are Riemannian indices in ten dimensions,
$A=(a,m)$ are local Lorentz indices,  $a,\alpha = 0,\cdots,4$, $m,
\xi =5,\cdots,9$; $\gamma^a$ and $\tau^m$ are the Dirac matrices in
$AdS_5$ space and $T^{1,1}$, respectively;  $e^A_{~~M}$ are vielbein
in ten dimensions,
\begin{eqnarray}
&& \left\{\Gamma^A,\Gamma^B \right\}=2\eta^{AB},
~~\left\{\Gamma^a,\Gamma^b
\right\}=2\eta^{ab},~~\left\{\tau^m,\tau^n
\right\}=2\delta^{mn},\nonumber\\
&& \gamma^0=-\gamma_0=i\sigma^1 \otimes 1_2=i
\left(\begin{array}{cc} 0 & 1_2 \\ 1_2 & 0 \end{array}\right), ~~
\gamma^i=\sigma^2 \otimes \sigma^i =\left(\begin{array}{cc} 0 &
i\sigma^i
\\ -i\sigma^i & 0 \end{array}\right),\nonumber\\
&& \gamma^4=\gamma_4= i\gamma^0\gamma^1\gamma^2\gamma^3=
\sigma^3\otimes 1_2=\left(\begin{array}{cc}  1_2 & 0 \\ 0 &-1_2
\end{array}\right), ~~~i=1,2,3,\nonumber\\
&& \tau^5=\tau_5=i\gamma_0,
~~\tau^6=\tau_6=\gamma_1,~~\tau^7=\tau_7=\gamma_2,
~~\tau^8=\tau_8=\gamma_3,~~\tau^9=\tau_9=\gamma_4.\nonumber\\
&& G_{MN}=\eta_{AB}e^A_{~~M}e^B_{~~N}, ~~~\eta_{AB}=\mbox{diag}(-1,
1, \cdots,1)
 \label{gmat2}
\end{eqnarray}
The above  choice on $\Gamma$-matrices determine the ten-dimensional
$\gamma_5$-analogue,
\begin{eqnarray}
\Gamma^{11}=\Gamma_{11}=\Gamma^0 \Gamma^1\cdots \Gamma^9=1_4\otimes
1_4\otimes \sigma_3=\left(\begin{array}{cc} 1_{16} & 0 \\ 0 &
-1_{16}
\end{array}\right).
\end{eqnarray}

\end{document}